# Indications of stellar coronal mass ejections through coronal dimmings

Astrid M. Veronig[1,2]*, Petra Odert[1], Martin Leitzinger[1], Karin Dissauer[1,3], Nikolaus C. Fleck[1], Hugh S. Hudson[4,5]

1 Institute of Physics, University of Graz, Universitätsplatz 5, 8010 Graz, Austria
2 Kanzelhöhe Observatory for Solar and Environmental Research, University of Graz, Kanzelhöhe 19, 9521 Treffen, Austria
3 NorthWest Research Associates, 3380 Mitchell Lane, Boulder, CO 80301, USA
4 School of Physics & Astronomy, University of Glasgow, Glasgow G12 8QQ, UK
5 Space Sciences Laboratory, University of California, Berkeley, CA 94720, USA
*Correspondence to: astrid.veronig@uni-graz.at

**Coronal mass ejections (CMEs) are huge expulsions of magnetized matter from the Sun and stars, traversing space with speeds of millions of kilometers per hour. Solar CMEs can cause severe space weather disturbances and consumer power outages on Earth, whereas stellar CMEs may even pose a hazard to the habitability of exoplanets. While CMEs ejected by our Sun can be directly imaged by white-light coronagraphs, for stars this is not possible. So far, only a few candidates for stellar CME detections are reported. Here we demonstrate a different approach, based on sudden dimmings in the extreme-ultraviolet (EUV) and X-ray emission caused by the CME mass loss. We report dimming detections associated with flares on cool stars, indicative of stellar CMEs and benchmarked by Sun-as-a-star EUV measurements. This study paves the way for comprehensive detections and characterizations of CMEs on stars, important for planetary habitability and stellar evolution.**

CMEs expelled from the Sun to interplanetary space are frequently observed, with occurrence rates of about five per day during solar maximum and one per day during solar minimum periods. Fast CMEs are generally associated with flares, i.e. sudden outbursts of radiation. Flares and CMEs are different facets of a joint underlying physical process that involves instabilities and reconnection of magnetic fields in active regions[1,2]. Late-type stars have outer convection zones, thus exhibiting stellar activity and maintaining coronae similar to our Sun. They produce flares which can be orders of magnitude more energetic than solar flares[3]. Stellar flares are common, observed in particular at soft X-ray and optical wavelengths[4,5]. The situation is very different for the observations of stellar CMEs. While solar CMEs can be directly imaged by coronagraph instruments blocking the million-times-brighter radiation from the Sun's surface, and their speeds and masses derived thereby, for stellar CMEs this is not possible. However, stellar CMEs and flares have recently gained increased attention, due to their potential impact on stellar evolution[3,6] and the habitability of exoplanets[7,8]. Further relevance is given by the recently detected "superflares" on solar-like stars[9], and their implications for the most disastrous space weather events our Sun might produce.

Different approaches have been explored in attempts to observe signatures of stellar CMEs. So far, the most promising one is the Doppler signal produced by CME-related plasma motions detected on a few main- and pre-main-sequence M-stars in optical[10-15] and UV spectra[16,17] as well as one X-ray detection on a giant star[18]. Searches for type II radio bursts generated by shock fronts driven by fast CMEs have not yet succeeded[19,20]. Indirect CME signatures have been reported in the form of increased X-ray column densities and transient UV absorptions during flares[21-23]. In total, the reported stellar CME detections are less than 20, and they are not unambiguous[18,24,25]. Here we present a different and more direct approach, the study of sudden decreases of the EUV and X-ray emission from solar and stellar coronae due to the matter ejected by CMEs, so-called coronal dimmings.

Coronal dimmings caused by CMEs have been regularly observed on the Sun for more than 20 years at EUV and soft X-ray wavelengths[26-28]. They are observed over a broad range of coronal temperatures, but show up most prominently for the 1-2 Million Kelvin plasma of the quiet corona[29,30]. Dimming regions map to the bipolar ends of closed magnetic field lines that become stretched or temporarily opened during an eruption, and are a result of the depletion of coronal plasma caused by the expansion and mass loss due to the CME[26,27]. Recently available multi-point imagery from satellites at different locations in the heliosphere provided us with unprecedented “quadrature” observations of the three-dimensional evolution of solar CMEs and their associated coronal dimmings. For a unique set of 76 events, coronal dimmings were observed on-disk by the Atmospheric Imaging Assembly (AIA)[31] onboard the Solar Dynamics Observatory (SDO), while the CMEs causing them were observed close to the solar limb by the Solar TErrestrial RElations Observatory (STEREO)[32] minimizing projection effects in the determination of CME speeds and masses[30,33]. These studies showed that in narrow-band EUV imagery, the spatially resolved dimming regions reveal emission decreases of up to 70%[30]. Distinct correlations were found between CME mass and speed, with key parameters of the associated coronal dimmings such as spatial extent and intensity drop[33,34]. Notably, CME-associated dimmings have been also identified in various ionization stages of iron lines (Fe IX to Fe XIV) in full-Sun EUV irradiance measurements[35,36,37].

Here, we evaluate Sun-as-a-star EUV irradiance measurements by SDO’s Extreme ultraviolet Variability Experiment (EVE)[38] as a testbed to study whether coronal dimmings can be also observed on stars and used for stellar CME detection. To this aim, we derive broad-band Sun-as-a-star EUV light curves from the SDO/EVE spectra for a set of large flares, analyse these light curves for the occurrence of significant coronal dimmings, and evaluate the association between coronal dimmings and CMEs (cf. Methods M1). This thorough approach allows us to quantify i) how frequently coronal dimmings are observed in Sun-as-a-star EUV light-curves as a signature of the accompanying CME, ii) how robust the CME detection via dimmings is against false alerts, and iii) how well do coronal dimmings serve as a proxy for CMEs. The data set consists of 44 large flares (GOES class ≥M5) observed in SDO/EVE irradiance spectra. The events have been classified in previous studies[37,39] into eruptive (i.e. associated with a CME: 38) and confined (i.e. no accompanying CME: 6) flares. Table 1 shows the contingency table for the CME and dimming occurrences, from which we infer a high conditional probability for the occurrence of a CME given that a coronal dimming was observed in the 15-25 nm broad-band

SDO/EVE light curve in the aftermath of a large flare, P(CME | Dim) = 0.970. In addition, we infer from Table 1 that coronal dimmings are a frequent phenomenon associated with CMEs, P(Dim | CME) = 0.842, whereas the probability of false alerts is small, P(Dim | !CME) = 0.167. These findings demonstrate that coronal dimmings caused by CMEs have been identified in spatially unresolved broad-band EUV data, with the emission dropping by 0.4% to 5.5% relative to the pre-event level (Supplementary Table 1).

| **EVE** | | !Dimming | Dimming |
|---|---|---|---|
| | !CME | 5 | 1 |
| | CME | 6 | 32 |
| **AIA** | | | |
| | !CME | 14 | 2 |
| | CME | 13 | 39 |

**Table 1:** Contingency table for the occurrence of CMEs and coronal dimmings in association with large flares (GOES class ≥M5), as derived from SDO/EVE 15-25 nm broad-band light curves (top) and full-disk integrated SDO/AIA 19.3 nm light curves (bottom).

To increase statistics, the analysis has been expanded by integrating the spatially resolved emission from SDO/AIA 19.3 nm full-disk images to create full-Sun light curves for a larger set of 68 flares (52 eruptive, 16 confined)[37,39]. The probabilities derived from this extended data set give a consistent picture with those derived from SDO/EVE, with P(CME | Dim) = 0.951, P(Dim | CME) = 0.750, and P (Dim | !CME) = 0.125. Note that in the subsample overlapping with SDO/EVE, for 42 out of 44 events the EVE and AIA dimming identifications agree. These findings provide strong evidence that dimmings are a good and robust proxy for CMEs, since i) with very high probability the coronal dimmings identified in the EUV light curves are due to a CME accompanying the flare, and ii) CMEs associated with large flares frequently manifest themselves as significant coronal dimmings in Sun-as-a-star EUV light curves. Under the plausible assumption that the coronal environment in other late-type stars behaves in a similar way than our Sun, coronal dimmings are an excellent means for stellar CME detection.

Fig. 1 and the accompanying Supplementary Video 1 demonstrate for the March 7, 2012 event the relation between the spatially resolved dimming in SDO/AIA and the unresolved Sun-as-a-star observations by SDO/EVE, together with its CME. This event reveals the strongest EVE dimming, and is caused by a very fast and massive CME (maximum speed of 3700 km/s and mass of $1.8\times10^{16}$ g)[33], followed by another fast CME originating from the same active region within 1 hour. AIA 19.3 nm images (Fig. 1, first column) show the bright solar flare as well as localized dark dimming regions on both ends of the flare arcade. Corresponding base-ratio images (Fig. 1, second column), where each frame was divided by a pre-event image to enhance changing features, reveal the global nature of the dimming. The third column in Fig. 1 shows the double CME that caused this dimming, as observed by EUVI and COR2[32] onboard STEREO-B located 118° East of Earth.

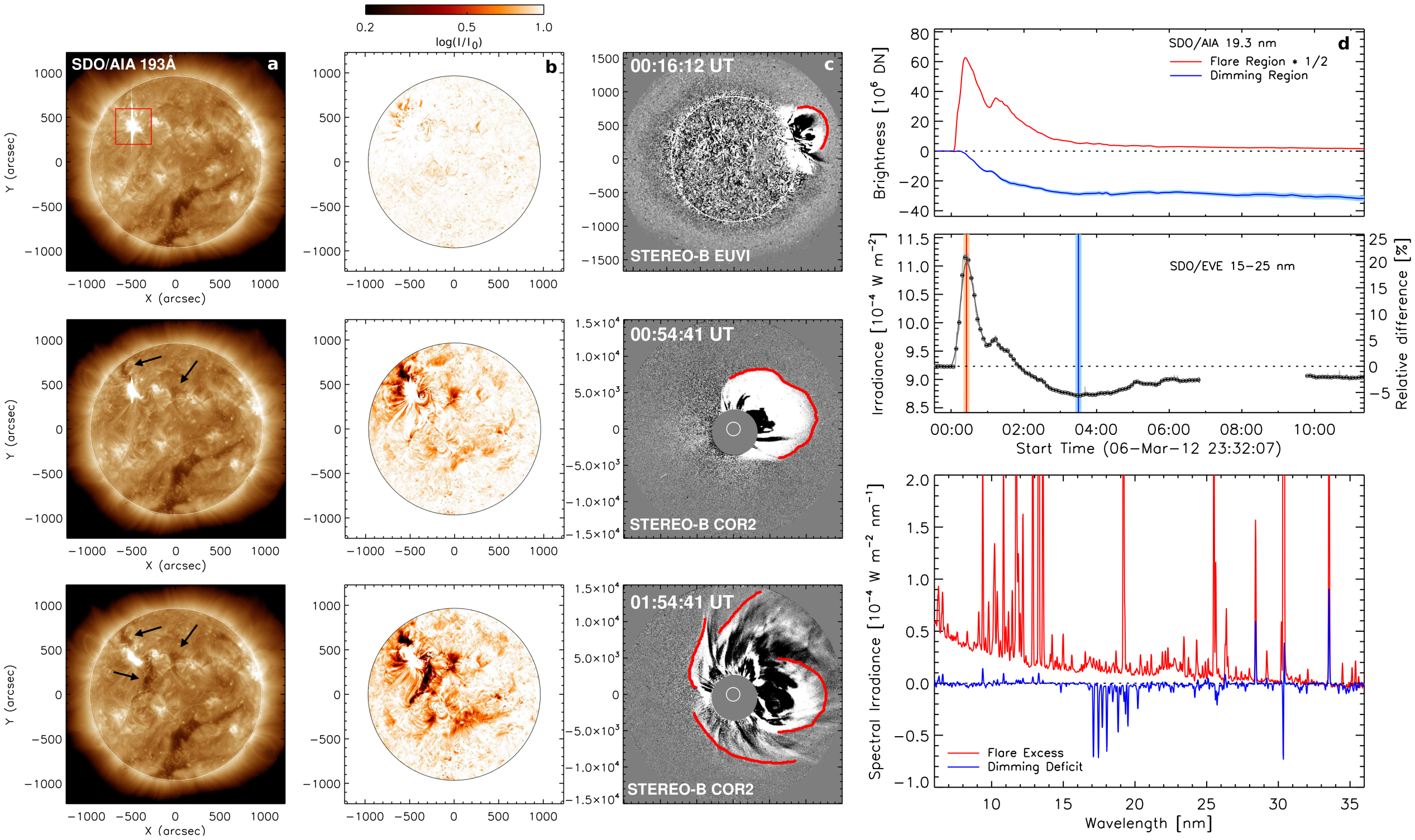

**Figure 1: Coronal dimming event on Sun on March 7, 2012**. Columns a,b): SDO/AIA 19.3 nm direct and logarithmic base-ratio images showing the flare and coronal dimming. The color bar on top marks the logarithmic values (dimensionless) indicative of the changes with respect to the pre-event emission for the images shown in column b. c) CME imaged by STEREO-B EUVI and COR2. d) Top: Spatially resolved SDO/AIA 19.3 nm light curves of the flare (red) and dimming region (blue); pre-event emission is subtracted. Middle: SDO/EVE 15-25 nm broad-band Sun-as-a-Star light curve. Bottom: pre-event subtracted SDO/EVE irradiance spectra, integrated over 10-min during the flare peak (red; the plotted y-range is cut) and over the maximum dimming depth (blue), as indicated by vertical lines in the middle panel. Supplementary Video 1 is available with this figure.

The fourth column of Fig. 1 shows the relation between spatially resolved and unresolved EUV light curves, which is key to make the transition to the stellar case. To understand the interplay of two competing effects in the spatially unresolved SDO/EVE light curves, i.e. the emission enhancement due to the flare and the decrease due to the dimming, we use SDO/AIA imagery to derive light curves separately for the flare and the dimming regions. The flare emission increase and the dimming emission decrease start roughly simultaneously, which can be understood in terms of the coupling of the CME dynamics and the flare energy release by magnetic reconnection in the large-scale current sheet beneath the erupting structure[1,2]. However, the dimming continues much longer, enabling its detection in the unresolved data. The flare duration is related to the energy release and cooling time scales of the hot flaring plasma, whereas the dimming duration is related to the time scales for replenishment of the solar corona in the aftermath of a CME and the related mass loss. The evolution of the segmented components is reflected in the SDO/EVE 15-25 nm Sun-as-a-star light curves, which are initially dominated by the strong flare enhancement, but after about 110 min the emission decrease due to the dimming starts to prevail. The bottom right panel shows two SDO/EVE spectra, subtracted by a pre-event spectrum: the red one is centered at the peak of the flare, the blue one at the maximum depth of the dimming. The spectra show strongly enhanced flare continuum and spectral lines, whereas the dimming is dominated by spectral lines, most prominent in the wavelength range 17-21 nm.

In Extended Data Fig. 1, we show broad-band EUV and X-ray full-Sun light curves of five more examples of CME-associated flares to illustrate various dimming cases: a) dimming interrupted by a subsequent large flare, b) impulsive flare where the dimming shows up promptly, c) long-duration flare followed by a very weak but significant dimming, d) very gradual dimming, e) strong dimming that remains at deep level for >11 hrs. Extended Data Fig. 2 shows for comparison a) an example of a confined flare with no dimming, and b) the only confined flare where a significant EVE dimming was identified, albeit much weaker and less impulsive than most of the dimmings associated with eruptions.

As we have shown here that coronal dimmings are detectable in broad-band Sun-as-a-star EUV light curves, with a high conditional probability that the occurrence of a dimming in association with a large flare is accompanied by a CME, P(CME | Dim) = 0.97, we can apply these important findings to the stellar case. We focus on Sun-like and late-type (F, G, K, M) main-sequence and pre-main sequence stars. Coronal temperatures of active Sun-like stars are correlated with their X-ray emission[40], and more active stars may reveal higher coronal temperatures than the Sun[41]. This implies that stellar dimmings may be also observable at shorter wavelengths. Thus, we explored data archives of missions operating at EUV (NASA's Extreme Ultra-Violet Explorer/EUVE[42], 7-76 nm) as well as soft X-ray wavelengths (ESA's X-ray Multi-Mirror Mission/XMM-Newton[43], 0.1-6 nm; NASA's Chandra[44], ~0.1-20.7 nm), compiling observations from 201 stars (cf. Methods M3). In order to identify flares, we constructed light curves covering also the pre- and post-flare phases. These were subsequently analysed for flare-associated dimmings indicative of stellar CMEs, applying the following criteria: a) after the flare, the EUV or X-ray flux drops significantly ($>2\sigma$) below the quiescent pre-flare level in at least one energy band, and b) after reaching a minimum, the flux increases again towards the pre-flare level (cf. Methods M5, M6). This analysis yields 21 dimmings detected on 13 different stars, one from EUVE, 3 from Chandra, and 17 from XMM-Newton (Supplementary Table 2). Notably, about half of the events are found on three stars: the young and rapidly rotating K0V star AB Dor (5 events), the young M0Ve star AU Mic (3 events) and the nearby M5.5Ve star Proxima Centauri (2 events). The other events were found on G- to M-type pre-main sequence and main-sequence stars.

Figure 2 shows two dimming events detected on Proxima Centauri. The XMM 0.2-2 keV light curve (panel a) shows that the dimming starts immediately after a small, short flare, reaches a maximum depth of 36% and lasts for approximately three hours before it is interrupted by three subsequent flares. During the indicated dimming interval, also one small flare occurs. Panel c shows a small flare followed by a pronounced dimming with a decrease of 56% and a duration of 4.5 hours (again interrupted by subsequent flares). All flares are also seen in the photometric U-band, revealing a very stable flux level outside flaring periods (panels b, d). In the bottom panels, we show spectra as well as temperature and emission measure evolution derived from spectral fits. They reveal a strong emission measure drop during the dimming, providing further support that the stellar dimmings are due to coronal mass loss.

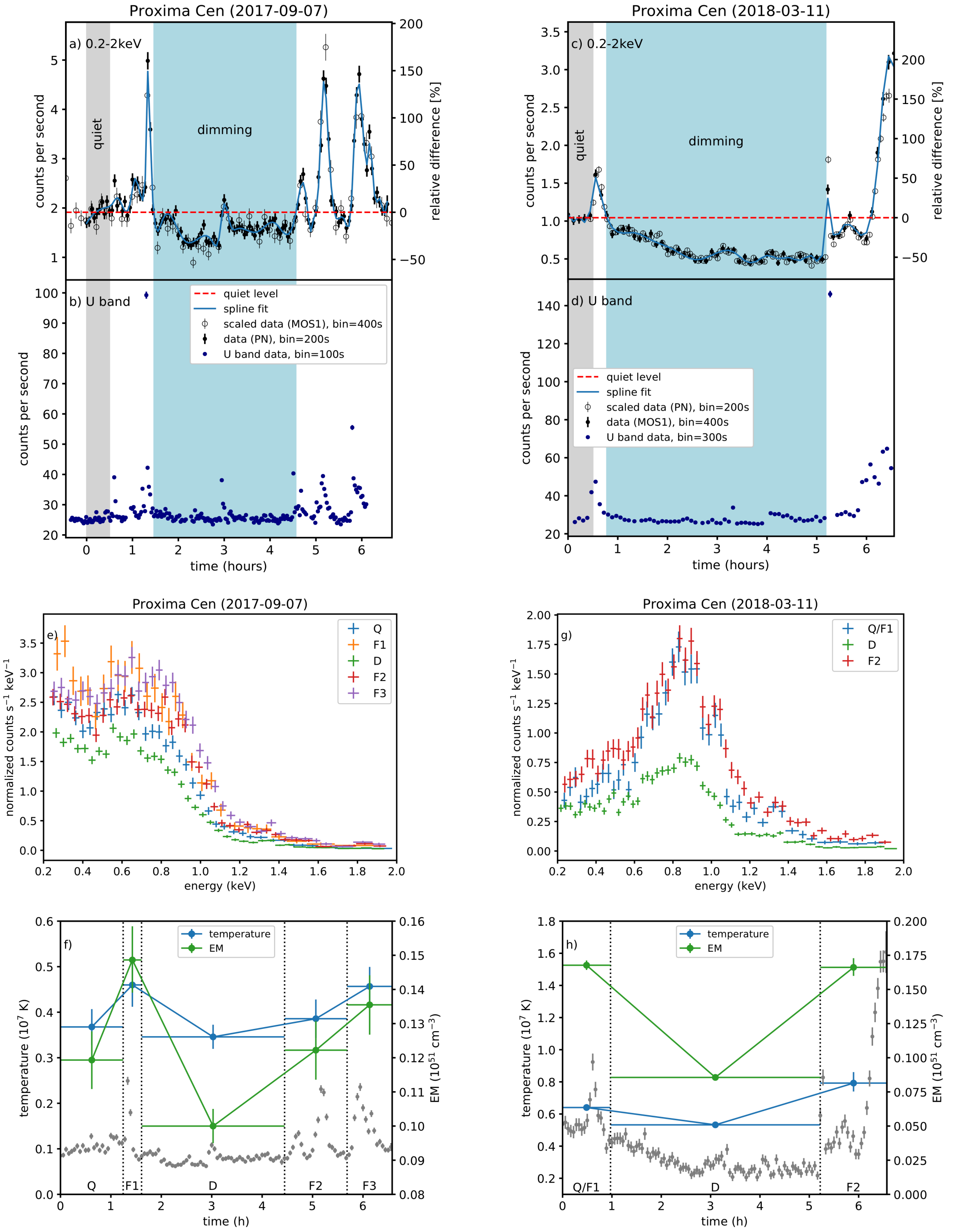

**Figure 2: Coronal dimmings on Proxima Centauri.** First row: Background subtracted XMM X-ray (0.2-2 keV) light curves. The weighted spline fits (blue) to the PN (a) and MOS1 (c) data as well as the adopted quiet levels (red) are shown. Dimming and quiet intervals are highlighted by blue and grey shaded areas. Second row: Simultaneous photometric fast mode OM observations in the U-band. Third row: PN (e) and MOS1 (g) spectra in the energy range 0.2-2 keV for different time intervals. Fourth row: Total emission measure and emission-measure-weighted temperature from spectral fitting. The time intervals are indicated by vertical lines, and the light curves (arbitrarily scaled) are shown in grey. Error bars are 1-sigma uncertainties.

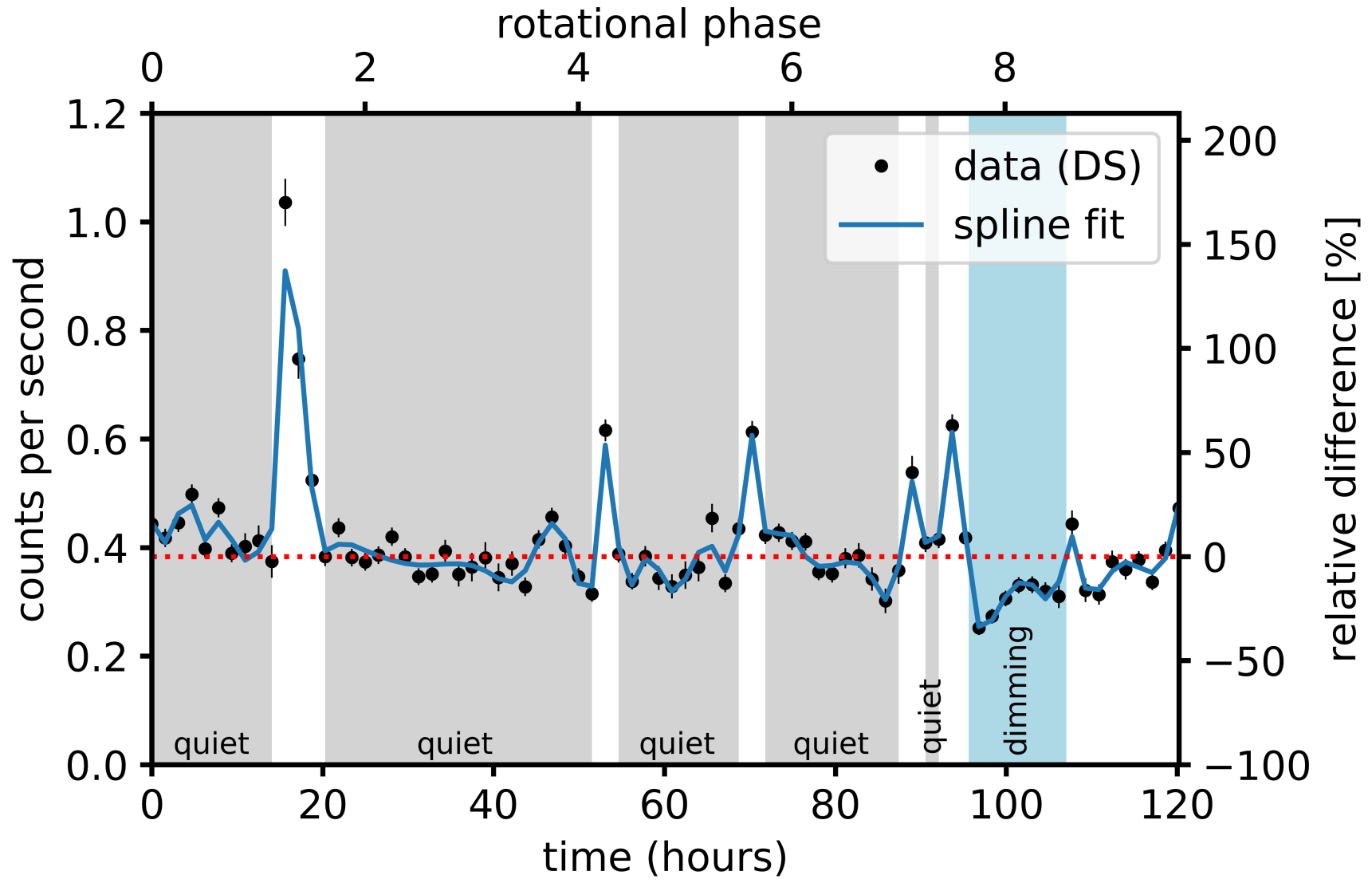

**Figure 3: Coronal dimming on AB Dor.** Background subtracted EUVE DS (8-18 nm) light curve with data binned to 5620 seconds corresponding to one EUVE orbit. Error bars give 1-sigma uncertainties. The blue line shows the weighted cubic spline fit. Grey shaded areas denote periods used to determine the quiescent level (red dashed line). The light-blue area marks the dimming.

Figure 3 shows the only dimming event detected with EUVE, in its DS (8-18 nm) data. The light curve shows five flares. After the last one at 93.7h, the count rate drops below the quiescent level by 34% followed by a gradual recovery, with a total dimming duration of 11.4 hours. The data shown cover ~9 rotations of AB Dor. They reveal a stable quiescent level and no obvious rotational modulation[45]. In total, we identified five dimming events on AB Dor (one in EUVE, four in XMM data; cf. Extended Data Figs. 3 and 4). An important issue for dimmings occurring on rapidly rotating stars is the maximum observability of the source region. If the star has an inclination $i$ = 90°, an active region can generally be detected for half of a stellar rotation period until it rotates off the disk. AB Dor has $i$ = 60°,[46] which means we can see the star's pole. Doppler imaging has revealed a large polar starspot on AB Dor[47,48]. In Fig. 3, the duration from flare peak to dimming end is 13.3 hours, which is more than one rotation period. This observability over a full rotation suggests that the observed flare and related CME/dimming originate from a starspot located in a polar region. A very recent study[49] based on TESS data suggests that ~60% of all flares on AB Dor occur on its constantly visible polar region, which strongly supports our findings. Light curves for all stellar dimming events are shown in the Extended Data Figs. 3 to 10.

From the stellar light curves, we derived characteristic dimming parameters: a) maximum depth, b) duration, c) rise time from dimming start to maximum depth, d) recovery time from dimming maximum back to quiet level, e) delay between flare peak and dimming maximum, f) delay between flare peak and dimming start (cf. Methods M6). The parameters of all the stellar dimmings identified and information on the host stars are listed in Supplementary Table 2. The distributions of these parameters are shown in Figure 4, along with the same plots for the solar dimmings identified in the SDO/EVE 15-25 nm light curves (cf. Methods M1). The stellar dimming depths range from 5% to 56%, which is an order of magnitude larger than the solar dimmings.

This difference can be explained as an observational selection effect: the quiescent levels of stellar light curves show much larger variances, thus only strong dimmings can be identified. The dimming rise times are similar in both cases, with a mean of ~2 hours. Due to subsequent flares or the end of observations, the duration and recovery times represent a lower limit in most of the stellar dimmings and in at least 17 of the solar dimmings. The time between the flare peak and the dimming start tends to be longer for the stellar (mean of 2.3 hours) than the solar dimmings (0.9 hours). We assume that the larger delay is due to the stronger emission and longer decay times in stellar flares, thus obscuring the effect of the dimming for a longer period (cf. Figure 1).

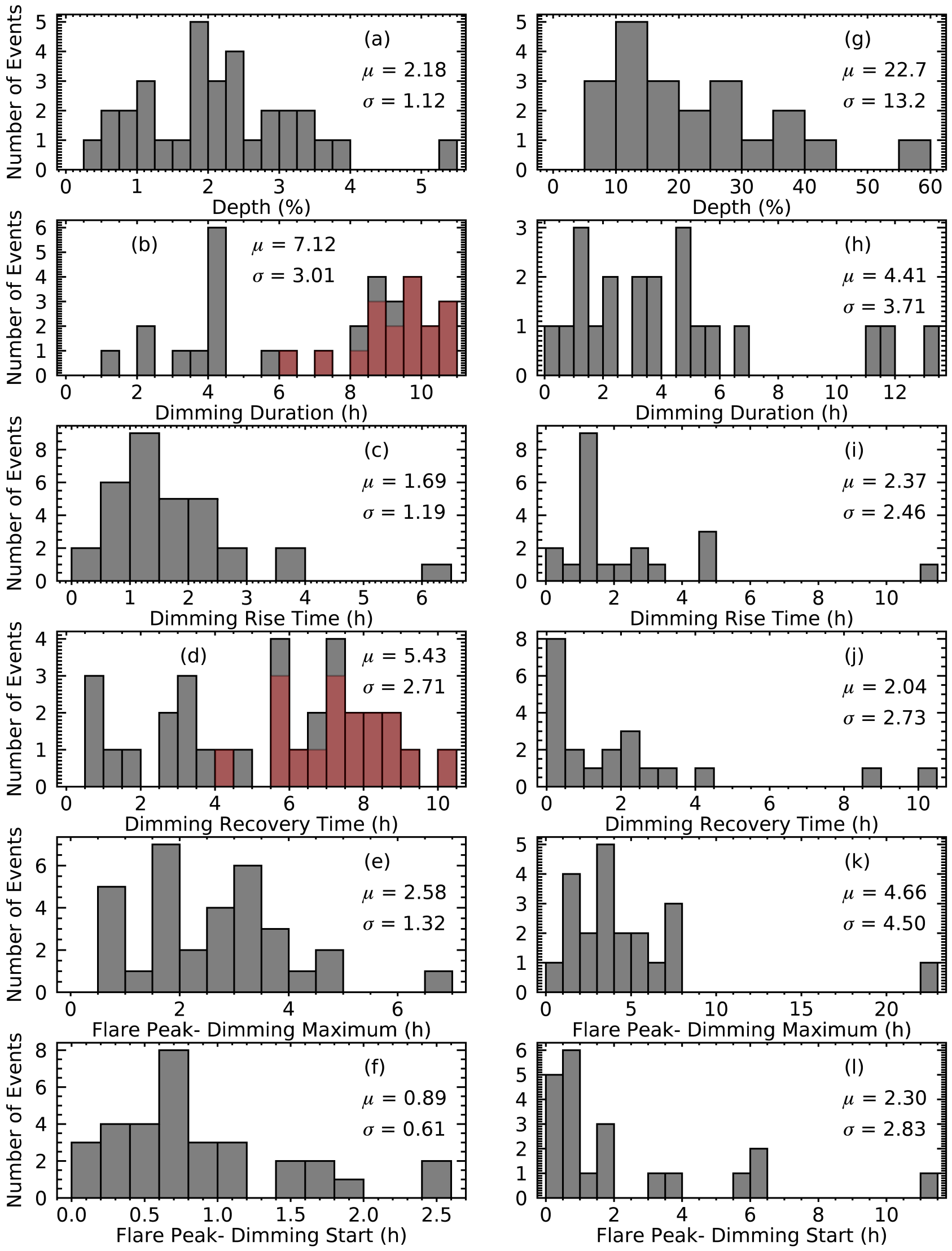

**Figure 4: Characteristics of solar and stellar dimmings.** Distributions of dimming parameters derived from the solar (left) and stellar (right) samples, with mean and standard deviation annotated. From top to bottom: dimming depth, duration, rise time, recovery time, delay between flare peak and dimming maximum, delay between flare peak and dimming start. Red-colored dimming duration and recovery times indicate solar events where the dimming end was not reached within the 12 hours observation interval.

This research has produced systematic detections of coronal dimmings on Sun-like and late-type flaring stars, suggestive of flare-related CME occurrence and mass loss. In total, we identified 21 CME candidates on 13 different stars, which is larger than the total number of all previous stellar CME detections reported[10-25]. The accordance of the characteristics of the stellar dimmings with solar dimmings identified in Sun-as-a-star broadband EUV data, their relation to the spatially resolved EUV imagery as well as the high association between dimmings and CMEs established for the Sun, strongly supports the interpretation of the stellar dimmings being caused by CMEs. The flux decreases due to the stellar CMEs in the range of 5–56% of the pre-event coronal emission are an order of magnitude larger than for solar dimmings. Depending on the emission mechanism dominating the dimming (lines or continuum bremsstrahlung), these findings suggest that the strongest stellar dimmings correspond to as much as half of the mass of the visible star's corona being depleted by the CME.

Our study provides the foundation for a different approach of detecting CMEs on stars, which may be extended toward estimations of their masses and speeds. These are key parameters for a better characterization of the contribution of CMEs to the mass and angular momentum loss of stars as well as the atmospheric escape and habitability of exoplanets[6,7,8]. The information derived will be maximized by combining dimming observations with physics-based simulations of stellar eruptions. A recent proof-of-concept study showed that magneto-hydrodynamic simulations have the potential for quantitative studies of stellar CMEs based on their coronal dimmings[50]. Interestingly, increasing the magnetic flux in the erupting flux-rope in these simulations, in order to make the step from the solar to the more energetic stellar case, suggested that stellar dimmings occur at higher temperatures of about 5 million Kelvin. This implies that stellar dimmings should be well visible in the EUV but also the X-ray domain, in line with the stellar dimming observations presented here. Future missions at EUV and X-ray wavelengths with higher sensitivity are expected to provide us with many more coronal dimming observations, to be used for stellar CME detections and characterizations. The dimming approach would be further strengthened by coordinated observations in other wavelength regions, such as optical spectroscopy, to obtain complementary information and to identify associated mass flows.

**Acknowledgments:**
We have made use of data obtained through the XMM-Newton Science Data Archive, operated by ESA at VILSPA, the Chandra Data Archive, operated by the Smithsonian Astrophysical Observatory for NASA, and the Mikulski Archive for Space Telescope for EUVE operated by the Space Telescope Science Institute. SDO data are courtesy of NASA/SDO and the AIA and EVE science teams. This research has made use of the SIMBAD database, operated at CDS, Strasbourg, France. The authors thank the members of the International Space Science institute (ISSI) team on “Coronal Dimmings and their Relevance to the Physics of Solar and Stellar Coronal Mass Ejections” for fruitful discussions. We thank J. Sanz-Forcada for help in the reduction of EUVE data. **Funding:** A.M.V., P.O., M.L., K.D. and N.C.F. acknowledge the Austrian Space Applications Programme of the Austrian Research Promotion Agency FFG (ASAP-14 865972, BMVIT). M.L. and P.O. acknowledge the Austrian Science Fund (FWF): P30949-N36.

**Author contributions:**
A.M.V. conceptualized and led the study, co-designed its solar part, and led the writing. P.O. co-designed the stellar part of the study and analysed the XMM-Newton data. M.L. co-designed the stellar part of the study and analysed the Chandra and EUVE data. K.D. co-designed the solar part of the study and analysed the SDO/AIA data. N.C.F. analysed the SDO/EVE data. H.S.H. contributed to SDO/EVE spectral analysis. All authors discussed the results and contributed to the manuscript text. A.M.V., P.O., M.L., K.D. and N.C.F. contributed equally to this work.

**Competing interests:**
The authors declare no competing interests.

**Additional information:**
**Extended Data** is available for this paper:
**Supplementary Information** is available for this paper:
**Correspondence and requests for materials** should be addressed to Astrid M. Veronig (astrid.veronig@uni-graz.at).

## Methods

### M1. SDO/EVE Sun-as-a-star solar dimming identification and CME association

SDO/EVE measures the spatially unresolved solar spectral irradiance (Sun-as-a-star) over the wavelength range 6-105 nm. The main instrument is the Multiple EUV Grating Spectrograph (MEGS), consisting of a grazing-incidence spectrograph (MEGS-A, 6-37 nm) and a two-grating, cross-dispersing spectrograph (MEGS-B, 35-105 nm), both at a spectral resolution of 0.1 nm.[38] EVE Level-2 spectral data were used over the whole wavelength interval, with 0.02 nm spectral binning and 10 s cadence. From these Level-2 data, we created broad-band light curves by integrating the spectra over the wavelength range 15-25 nm, where coronal dimmings are well observed. For each event, we generated a light curve covering a duration of 12 hours, starting 60 min before the commencement of the associated GOES X-ray flare. To enhance the signal-to-noise level and for better comparison to the stellar observations, the data were rebinned to 5-min intervals. The pre-event quiet level was determined as the mean of the last nine 5-min data points before the flare start, and the corresponding standard deviation was used to give an estimate of the errors on the data points. For robustness, the dimming parameters (Figure 4a-f, Supplementary Table 1) were determined from the 15-25 nm light curves smoothed with a weighted spline fit.

Our solar data set is based on two extensive lists of large solar flares covering the full SDO/EVE era that have been classified in previous, independent studies into eruptive/confined events, i.e. into flares with/without an accompanying CME. The Baumgartner et al. (2018)[39] study comprises 44 flares of GOES class M5.0 and larger that took place within 50° from Sun center (32 eruptive, 12 confined) during 01/2011 to 12/2015. The Harra et al. (2016)[37] study covers 42 X-class flares (33 eruptive, 9 confined) without any constraints on the source location, i.e. also including flares on the limb, that occurred during 02/2011 and 11/2014. We note that the lists are partially overlapping, 17 events appear in both catalogs. The SDO/EVE MEGS-A spectrograph was operational from 30 April 2010 to 26 May 2014. 28 of the Baumgartner et al. (2018)[39] and 31 of the Harra et al. (2016)[37] events occurred during the period of EVE operations, with 11 events being included in both lists. This results in a total of 48 large flares. Four of the events had to be removed from further analysis due to one of the following reasons: incomplete EVE data coverage, too close succession of a major flare, or no clear pre-flare level being available. The final EVE data set for our Sun-as-a-star analysis thus comprises 44 large flares (38 eruptive, 6 confined). In order to increase the statistics, in particular for confined flares, we expanded the analysis by using also the SDO/AIA 19.3 nm filtergrams, which are centered at an EUV spectral line that is very prominent in dimmings[30] (see also the spectrum in Figure 1). We used AIA 19.3 nm image sequences with 5 min cadence, from which we created quasi-Sun-as-a-star light curves by summing for each time step the counts of all pixels of the image. In this way, we analysed all the events listed in the two catalogs[37,39] with the same methods as the SDO/EVE light curves, resulting in 68 large flares (52 eruptive, 16 confined).

In order to use solar dimmings as a benchmark to evaluate whether dimmings are a viable means for CME detection in solar-like and late-type stars, one needs to quantify several aspects: a) How large is the fraction of CMEs accompanying large solar flares that reveal a significant dimming in the EVE broad-band Sun-as-a-star light curves? If this is a large fraction, then by solar-stellar analogy, the dimming phenomenon has a high potential to show up also in stellar flares associated with CMEs. b) How large is the probability of false positives, i.e. cases where a significant dimming in the aftermath of a large flare is detected although there was no CME associated? It is important that this number be small compared to the number of real detections, to make sure that the identified dimmings associated with flares are likely due to a CME accompanying the flare and not caused by some other variability in the full-Sun light curves. c) How large is the conditional probability that, given a coronal dimming occurred in the aftermath of a large flare, also a CME occurred? If this number is high, it provides strong evidence that coronal dimmings are a good proxy for CMEs. For the determination of the significance of coronal dimmings detected in the SDO/EVE 15-25 nm light curves (as well as the AIA 19.3 nm integrated light curves), we used the identical approach as for the stellar case (cf. Sect. M6). The only difference is that for the solar case we demanded that the time between the flare peak in the EVE broadband light curve and the time where the emission dropped below the level that marks a significant dimming be $\leq$3 hours. This criterion was applied to ensure robust associations between the flare and dimming, and to have clear objective rules to determine the entries for the contingency table, whether a flare was followed by a significant dimming and whether this dimming relates to a CME. We note that this time frame allowed is substantially longer than the typical rise time to the dimming maximum in spatially resolved observations (<40 min in 90% of the cases[30]), in which the dimming tends to start simultaneously with the flare (cf. Figure 1).

The calculation of the characteristic dimming parameters, such as maximum emission drop relative to the pre-event level, dimming rise time, duration etc. was done in the same way as for the stellar light curves (cf. Sect. M6). For the solar case, we implemented an automatic algorithm with the following rules. The dimming start was defined by the time where the spline fit to the data drops below the pre-flare level. For the dimming maximum, we differentiated two cases: If the EUV flux returned to the pre-event level within the 12-hrs time span under study, the deepest minimum between the start and end of the dimming period was used to define the maximum dimming drop. In the other cases (17 out of 44; indicated in red color in the histograms in Fig. 4b,d), the dimming maximum was defined by the first significant global minimum of the spline fit to the data that was reached after the detected dimming start.

**M2. Calculation of spatially-resolved flare and dimming light curves from SDO/AIA**
To derive spatially resolved light curves of the flare and the coronal dimming regions shown in Figure 1, we used EUV imagery of the SDO/AIA 19.3 nm filter, centered at Fe XII (log T=6.2) and Fe XXIV (log T=7.2) emission lines[31]. To derive the dimming light curve, at each time step we segment the instantaneous dimming region using a logarithmic base-ratio detection algorithm[51]. In order to quantify the absolute changes, we sum up the corresponding pixel values subtracted by the pre-event counts. To derive the segmented flare light curve, we detect at each instant all pixels that have counts exceeding the pre-event level by 50%, and sum up the values of all those pixels subtracted by the pre-event counts. To reduce the effect of CCD blooming on the flare masks, only pixels that were identified as flare pixels over at least 3 min were considered. The error ranges plotted in the spatially resolved flare and dimming light curves in Fig. 1 are calculated by varying the thresholds by ±5%.

**M3. Sample selection of solar-like and late-type flaring stars**
We have searched data archives of missions operating at EUV and soft X-ray wavelengths for flares observed on Sun-like and late-type (F, G, K, M) main-sequence and pre-main sequence stars. Using the approach described below, we identified observations from 201 stars suitable for our study (XMM: 51, Chandra: 114, EUVE: 36).

**M3.1 XMM-Newton**
The XMM-Newton Observatory[43] provides observations with three X-ray imaging detectors (EPIC PN, MOS1, MOS2), a high-resolution spectrograph and an optical/UV monitor (OM). The EPIC imaging detectors operate in the energy range 0.2-12 keV (0.1-6 nm)[52,53]. We used the XMM-Newton flare catalog[4] as basis for selecting events suitable for our study, and visually inspected all ~100 light curves classified as flares. The selected events were required to fulfill the following criteria: full coverage (including the pre-flare and post-flare phases) of at least one flare event; sufficient duration of the flare observation and/or sufficiently high count rate. This last criterion removed events with only one time bin classified as flaring. We compiled the data of all stars exhibiting flare observations fulfilling these criteria, and removed all stars from the sample that are not of spectral types F, G, K, or M. Further we removed binaries, except those with either well separated components, or those consisting of a primary Sun-like component with a close M dwarf, brown dwarf, or white dwarf secondary, where it can be assumed that the X-ray emission is dominated by the primary. We removed all evolved stars, but kept most pre-main sequence stars except for classical T-Tauri stars which may show signatures of accretion events in their light curves. This approach resulted in 36 events on 30 stars from the XMM-Newton flare catalog[4]. Since the catalog only includes observations up to the year 2007, we also evaluated more recent observations of the selected stars, identifying 7 more observations with suitable flares. In addition, we searched for data of a sample of well-known active flare stars[54], identifying 11 more stars with 23 suitable observations. Lastly, we checked the XMM-Newton Serendipitous Source Catalogue (3XMM-DR8). By selecting variable sources with sufficiently high count rates and detected variability, we found 20 additional events from visual inspection of the automated light curve plots provided. All together, this gives 86 observations of 51 different stars that were selected for further analysis of flare-associated dimmings.

### M3.2 EUVE

EUVE[42] (1992-2001) carried 4 telescopes, the deep sky telescope (DS, 8-18 nm), the short (SW, 7-18 nm), medium (MW, 17-37 nm), and long-wavelength (LW, 30-75 nm) spectrometers. No EUVE flare catalogue exists but there are studies on the flare activity of late-type main sequence stars[55]. We used the category "late" in the Mikulski Archive for Space Telescope (MAST) search form, which yielded 79 stars (not excluding binaries) from which we extracted a final target list containing 36 late-type main-sequence stars (70 observations).

### M3.3 Chandra

The Chandra X-ray Observatory[44] carries four science instruments, the High Resolution Camera (HRC; 0.06-10 keV), the Advanced CCD Imaging Spectrometer (ACIS; 0.08-10 keV), the High Energy Transmission Grating Spectrometer (HETGS; 0.4-10.0 keV) and the Low Energy Transmission Grating Spectrometer (LETGS; 0.4-10.0 keV), the latter two operating with ACIS. LETGS operates also with HRC covering then an energy range from 0.07-10.0 keV. No Chandra flare catalogue exists. Therefore, we used the Chandra category search, selecting all instruments and using the category "Stars and WD". We restricted the search for exposure times >5000 seconds, to cover flares including pre-flare periods. This search yielded 147 stars, including 114 late-type pre- and main-sequence stars (195 observations). We did not exclude binaries in this target sample.

## M4. Data preparation

### M4.1 XMM-Newton

For all events under study, we downloaded the *odf* files from the XMM-Newton Science Archive. Light curves were created using data from the three EPIC imaging detectors, using all three detectors in the initial analysis step. Usually all operate simultaneously, but in a few cases some detectors were running longer and may cover a part of the flare that is not included in the other exposures. If all exposure times were similar, we generally preferred the PN data due to their higher count rates. The XMM-Newton Scientific Analysis System (SAS) Version 17.0.0 was used for the data analysis. The SAS tasks *epproc* and *emproc* were used to create event lists from the downloaded *odf* files for the PN and MOS detectors, respectively. For observations included in 3XMM-DR8, we used the source and background extraction regions from the catalog. In all other cases, we manually selected circular regions from the EPIC images using *ds9*, where the background was placed on a source-free region, preferably on the same CCD.

For each event, we created light curves using the SAS task *evselect*, which was also used to create different energy binnings. Background subtraction and necessary corrections (vignetting, bad pixels, detector gaps, etc.) were carried out with the task *epiclccorr.* To check whether the considered observations are affected by pile-up, we used two approaches. First, we created diagnostic plots using the SAS tool *epatplot*. Since this only shows whether the whole observation is affected by pile-up, we also used the limiting fluxes from literature[56]. In most observations, only the flare peaks exceeded the pile-up limits, whereas the pre- and post-flare data were not affected. Therefore, in the further analysis we did not correct for pile-up, as dimmings are characterized by fluxes even below pre-flare levels. We note that although pile-up affects the different detectors differently, we found that the light curve morphology is similar in all available exposures with sufficient signal-to-noise. If available, the simultaneous optical OM data were reduced with the SAS tasks *omfchain* (fast mode) and *omichain* (imaging mode), respectively. We usually preferred the fast mode observations due to their higher time resolution, but used the imaging mode data if fast mode was not available.

**M4.2 EUVE**

EUVE data were retrieved via MAST, selecting night-time observations only (as daytime observations are contaminated by geocoronal emission), and reduced with the *euv1.9* package under IRAF 2.15. We used the X-ray PROS V2.5 package under IRAF 2.15, selecting the sources by circles with radii between 25-50 pixels and the background by annuli with radii between 25-100 pixels. For light curve extraction (background subtracted), we used the task *ltcurv* in the X-ray PROS package, in which we set the time binning as well as the task *effexp* in the EUV package accounting for instrument and telemetry dead times and vignetting.

**M4.3 Chandra**

We analysed Level-2 event files provided in the Chandra data archive using the Chandra Interactive Analysis Observations (CIAO) software package. For the non-grating observations, we used the task *dmextract* to extract background subtracted light curves. For the grating observations, we used the Interactive Spectral Interpretation System (ISIS) task *aglc*. For the Chandra grating data, we did not subtract a background because of low background levels. Data pile-up can have severe effects on the data analysis for Chandra ACIS data. For Chandra HETGS observations in higher orders (excluding 0th order), pile-up has only a small effect, as in the higher orders no high count rates are evident in single pixels. As all our events were detected in light curves constructed from higher order HETGS data, we did not apply pile-up correction.

**M5. Light curve binning**

For each event, we created light curves in the full available energy range as well as several subbands. Inspecting X-ray spectra of active Sun-like stars[57], for XMM and Chandra we select energy ranges in which strong spectral lines are formed at similar temperatures (0.6-1.2 nm ~ 1-2 keV: log T [K] ~ 7.0; 1.2-1.7 nm ~ 0.73-1 keV: log T [K] ~ 6.7; 1.7-6 nm ~ 0.2-0.73 keV: log T [K] ~ 6.3). This approach allowed us to study whether the identified dimmings occur at specific temperatures. All light curves were then inspected to select possible dimming candidates, defined to resemble characteristics of the solar observations. The flux was required to drop below the pre-flare level by at least $2\sigma$ (cf. Methods M6) in one or more energy bands after the flare, and to return back toward the pre-flare level after reaching a minimum. Events for which no sufficiently long (>0.5h) pre-flare level was available were removed from further analysis. In total, 21 stellar dimming candidates were identified fulfilling these criteria. In some cases, the observation ended before the flux returned to the pre-flare level. We derived the same set of dimming parameters as for the Sun (Supplementary Table 2). As stellar light curves are usually more noisy than solar observations, we determine the dimming parameters from weighted cubic spline fits to the flux curves.

The time binning was chosen individually based on the quality of the light curves. The obtained parameters slightly depend on the chosen time binning, but the spline fit is usually quite similar for time binnings that are not too coarse. To assess the significance of the detected dimmings, we need to obtain the uncertainty of the "quiet" flux level. This, however, depends strongly on the chosen time binning, as the typical errors on the count rate decrease with increasing bin size. Thus, the significance also increases with bin size. To obtain a more objective way of evaluating the uncertainties, we optimized the time binning for selected events by analysis of the Rayleigh periodogram of the raw event lists, where only the extraction of the source region and selection of the energy interval has been applied. By fitting the resulting Rayleigh periodograms with a power-law-plus-constant model, an optimal bin size for light curves can be obtained which best separates statistical noise from intrinsic variability[58]. We constructed this periodogram for all selected light curves between frequencies corresponding to the duration of the observation and the mean count rate. For fitting, we minimized the B-statistic using the Nelder-Mead algorithm. The optimal bin size was calculated as a function of the fit parameters[58]. We find that the light curves obtained with optimized binning and the spline fits to the data with smaller time binning are very similar in most cases, and the maximum dimming depths derived are consistent for both methods.

**M6. Determination of stellar dimming parameters and significance**

The determination of the dimming parameters depends on the choice of the quiet pre-flare level ($flux_{quiet}$), determined as the mean flux in the quiet time interval. This quiet interval was chosen immediately before the considered flare events, preferably covering a period >1h with low variability. Relative to this level, we define the start time of the dimming $t_{ds}$, where the weighted spline fit applied to the data drops below the quiet level for the first time after the flare. The minimum time $t_{dm}$, defined by the time of the first minimum of the spline fit ($flux_{dm}$) after the dimming start, is used to evaluate the maximum dimming depth as ($flux_{quiet}$ - $flux_{dm}$)/$flux_{quiet}$. The dimming end $t_{de}$ is defined as the first point after the dimming maximum where the spline fit returns back to the quiet level. Small flares that interrupt the dimming, but where the flux clearly drops below the quiet level afterwards again, are masked out from the dimming parameter determination. However, in all but three cases, either strong flares end the dimming, or the end of the observation is reached. Therefore, the calculated recovery times and durations of the stellar dimmings give lower limits.

To assess the significance of the dimming, we need to estimate the errors of the maximum dimming depths, which includes the mean error of the quiet flux and the mean error of the flux in the dimming region. The mean error of the quiet flux includes the contributions of both the error of the mean flux determined from the errors of the individual data points ($\sigma_i$, i = 1 … $N_{quiet}$) and the variance of the data points in the quiet interval, $\sigma_{quiet}^2 = \Sigma_i (\sigma_i/N_{quiet})^2 + var(flux_{quiet})$. This provides a maximum estimate of the mean error of the quiet level. The mean error of the dimming maximum ($flux_{dm}$) includes both the contributions from the spline fit (rms error) and the mean error of the flux in the dimming region ($t_{ds}$ to $t_{de}$), $\sigma_{dm}^2 = \Sigma_j(flux_{dm} - fit)^2/N_{dm} + \Sigma_j (\sigma_j/N_{dm})^2$ (j = 1 … $N_{dm}$), again providing a maximum error estimate. We consider the mean error averaged over the whole dimming region to be more representative than the error of a single data point at the maximum dimming depth. The error of the maximum dimming depth in flux units, depth = $flux_{quiet} - flux_{dm}$, is found by error propagation, $\sigma_{depth}^2 = \sigma_{quiet}^2 + \sigma_{dm}^2$. The error of the maximum dimming depth in percent is $\sigma_{\%}^2/depth_{\%}^2 = \sigma_{depth}^2/depth^2 + \sigma_{quiet}^2/flux_{quiet}^2$. We considered dimmings as significant if in at least one of the energy-binned light curves, the maximum dimming depth was larger than twice its error (>2$\sigma_{\%}$). Some dimming events may therefore be only significant in one energy band, and some are only significant if the optimal binning method is used. For the final results, we adopted the parameters derived from the spline fit method, because the timing parameters of the dimmings (duration, start, minimum, end time etc.) would have large uncertainties because of the much coarser time binning obtained with the optimized binning method. Also, with this method, the binning differs between different energy bands for the same event, making comparisons more difficult. The time of the flare maximum was taken from the original data, since the spline fit is not a good approximation of the sharp flare peaks. The stellar dimming parameters summarized in Supplementary Table 2 and Figure 4 were derived from the total energy band. Thus, not for all cases the dimming depth listed in column 7 of Supplementary Table 2 is >2$\sigma_{\%}$, as some dimmings are only significant in smaller subbands. The significance of the dimmings is indicated by different colors of the shadings in Extended Data Figures 3-9. We note that due to the long duration of the dimmings (typically >1h), the significance of the total flux deficit, i.e. over the whole dimming region, is much larger than that of the maximum dimming depth alone.

**M7. Spectral fitting**

For the two dimmings of Proxima Cen (Figure 2), we evaluate how plasma temperature and emission measure vary during the observation. We used the XSPEC software to fit spectra in selected time segments covering the whole light curves (each spectrum was required to have >3000 counts). The source+background and background spectra were created using the SAS task *evselect*, and *rmfgen* and *arfgen* were used to generate RMF and ARF files. We group the files with *specgroup*, which we also use to bin the spectra with a minimum of 20 counts per bin, while not oversampling the intrinsic resolution by a factor of 3. The model we chose includes an absorbing column (representing the ISM absorption; fixed to log $N_H$=17.61)[59] and several APEC plasma models representing the components of coronal plasma at different temperatures. The "quiet" pre-flare level was fitted first with one or more plasma components, where we checked whether adding further plasma components would significantly improve the fit. We then use this "quiet" solution as fixed input for the flare bins, but added further plasma components representing the flare. For the dimming regions, we only use the quiet solution as a starting value, and do not fix any parameters. We perform a simultaneous fit to all three detectors, but allow for a constant offset between detectors to account for uncertainties in absolute calibration[60]. We find that the quiet and dimming intervals are best fit with two plasma components, whereas the flaring intervals are best represented by three. We then calculate the total emission measure (EM= $\Sigma_i$ $EM_i$, for *i* plasma components) and the EM-weighted temperature (T= $\Sigma_i(T_i EM_i)$/EM).

**Data availability:**

The solar data used in this study are publicly available from the SDO/EVE data archive (http://lasp.colorado.edu/eve/data_access/eve_data/products/level2/) and the SDO/AIA data archive (http://jsoc.stanford.edu/ajax/lookdata.html?ds=aia.lev1_euv_12s). The stellar data used in this study are publicly available at the XMM-Newton Science Archive (http://nxsa.esac.esa.int/nxsa-web/#search), the Chandra Data Archive (https://cda.harvard.edu/chaser/) and the EUVE data archive at the Barbara A. Mikulski Archive for Space Telescopes (https://archive.stsci.edu/euve/search.php). The specific data sets used in this study are uniquely identified by their archive's Observation ID given in Supplementary Table 2. The resulting solar and stellar dimming parameters can be accessed through Supplementary Tables 1 and 2.

# Extended Data

## Indications of stellar coronal mass ejections through coronal dimmings

Astrid M. Veronig[1,2], Petra Odert[1], Martin Leitzinger[1], Karin Dissauer[1,3], Nikolaus C. Fleck[1], Hugh S. Hudson[4,5]

1 Institute of Physics, University of Graz, Universitätsplatz 5, 8010 Graz, Austria
2 Kanzelhöhe Observatory for Solar and Environmental Research, University of Graz, Kanzelhöhe 19, 9521 Treffen, Austria
3 NorthWest Research Associates, 3380 Mitchell Lane, Boulder, CO 80301, USA
4 School of Physics & Astronomy, University of Glasgow, Glasgow G12 8QQ, UK
5 Space Sciences Laboratory, University of California, Berkeley, CA 94720, USA

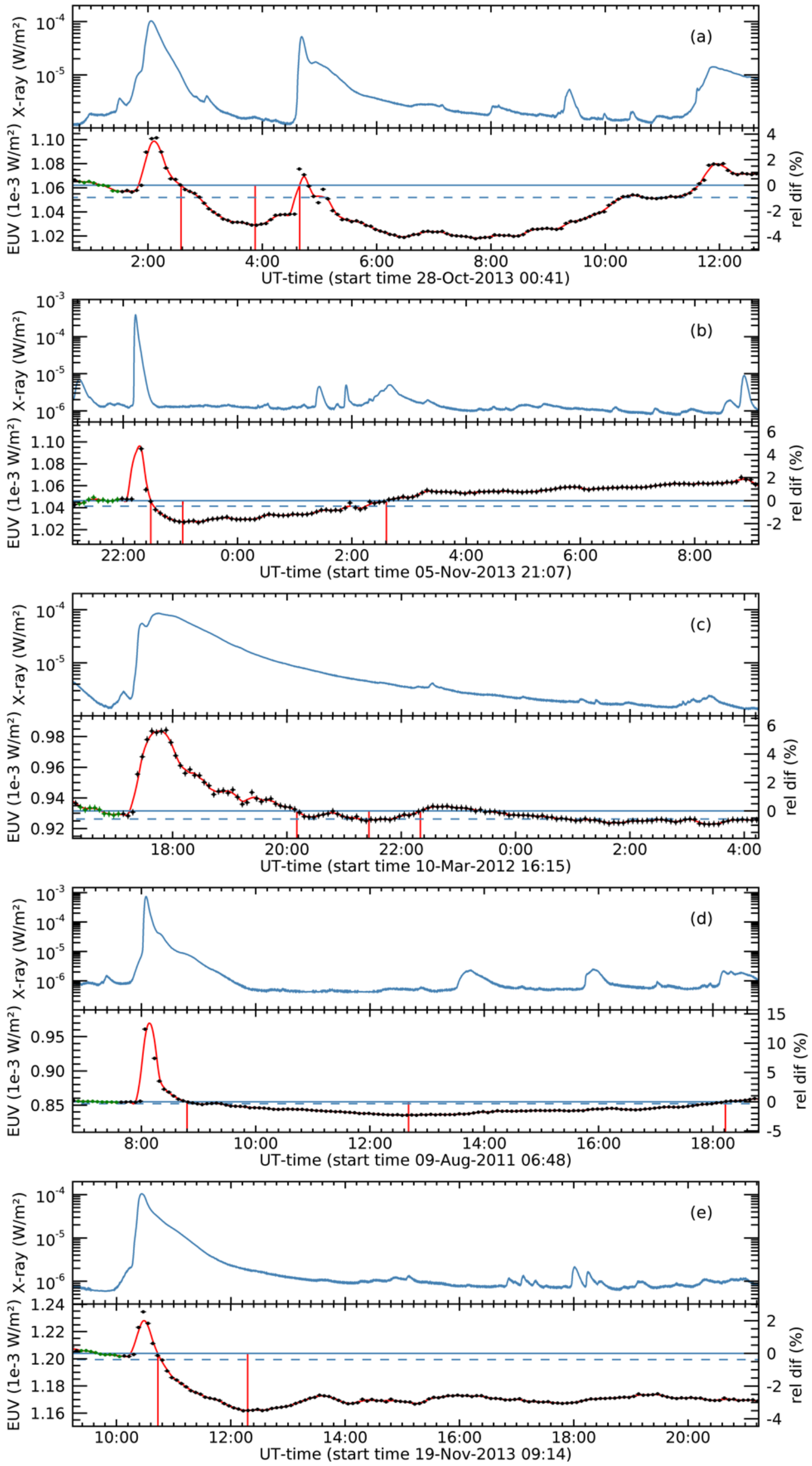

**Extended Data Figure 1**: Selected examples illustrating the manifold of coronal dimming appearances for CME-associated solar flares. Top subpanels: GOES full-Sun 0.1-0.8 nm soft X-ray flux, dominated by the hot flare emission. Bottom subpanels: SDO/EVE full-Sun 15-25 nm light curves showing the flares and coronal dimmings. The red line shows the weighted cubic spline fit. Green data points are used for calculation of pre-flare level. Horizontal lines represent the pre-flare (solid) and corresponding $2\sigma$ (dashed) level used to identify significant dimming emission decreases. Vertical lines indicate the times of the dimming start, maximum depth and end.

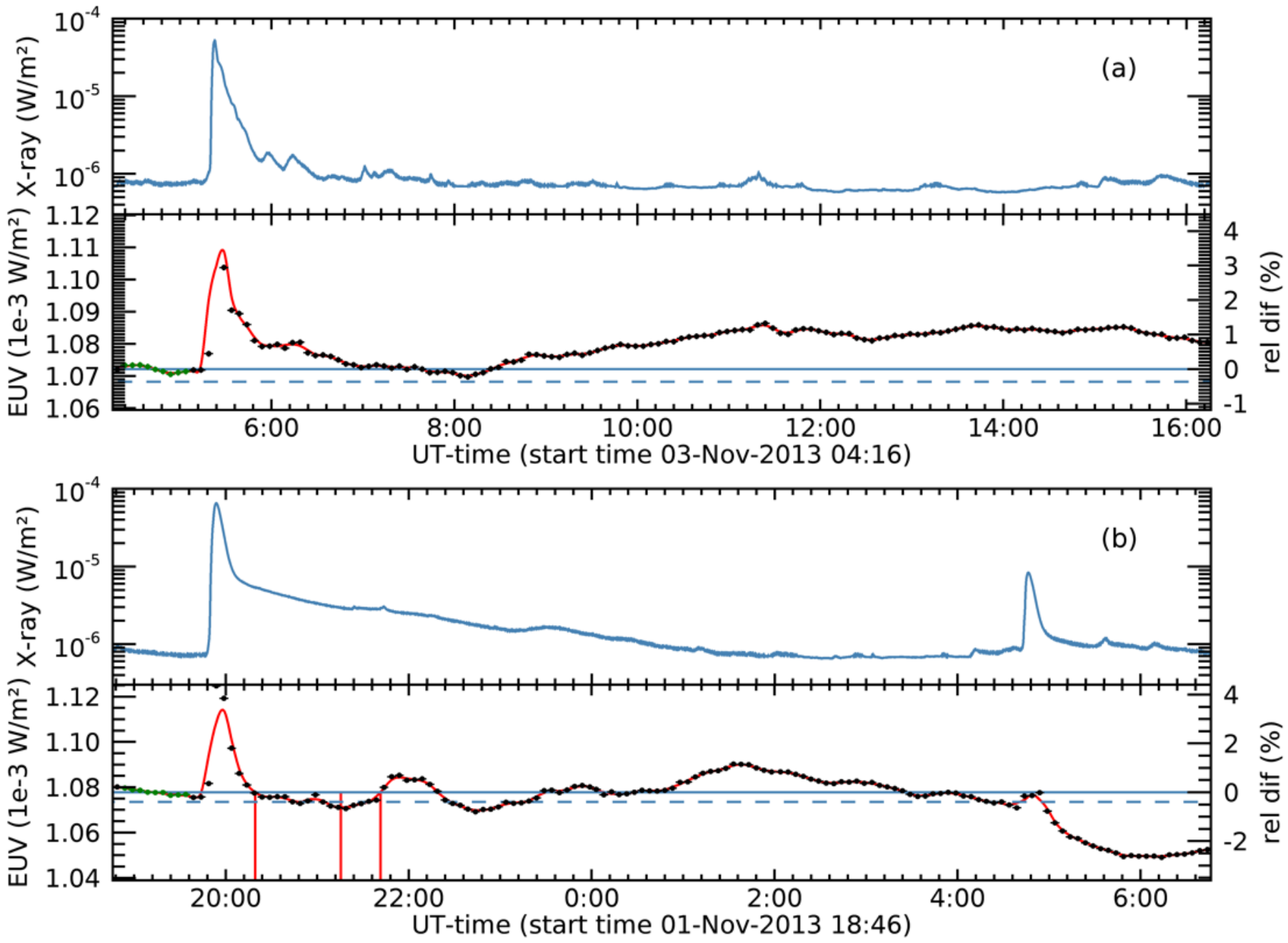

**Extended Data Figure 2**: Selected examples of light curves for confined solar flares. a) A confined flare (GOES class M4.9) which shows no dimming (true negative), b) the only confined flare (GOES class M6.3) where a significant EVE dimming was identified (false positive). Note that starting around 05:00 UT on 2 Nov 2013 there is a pronounced dimming associated with an eruptive C8.2 flare. Same plot format as Extended Data Fig. 1.

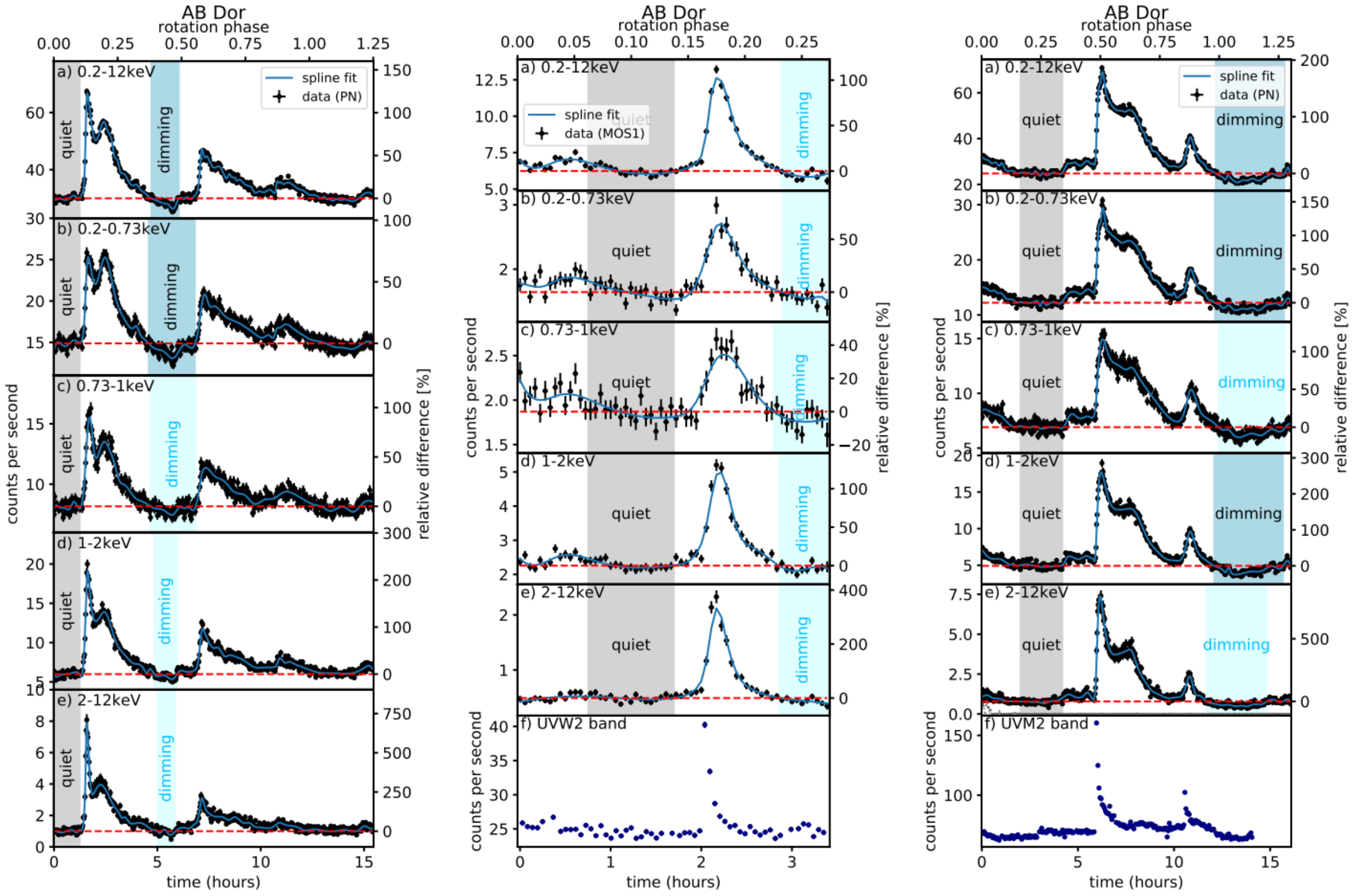

**Extended Data Figure 3:** Background subtracted X-ray light curves of the fast-rotating K-type star AB Dor. Data from the XMM EPIC PN (left and right columns) and MOS1 (middle column) detectors are shown as filled symbols, and are plotted for the total energy band (0.2-12 keV) as well as four subbands. Bin sizes of 100s, 200s, and 100s were used (from left to right). The weighted spline fits to the data are shown as blue lines, the adopted quiet levels as red dashed lines. The quiet time intervals are indicated as gray shaded areas, the detected dimming regions as blue shaded areas (dark blue: maximum dimming depth $>2\sigma_{\%}$, light blue: maximum dimming depth $<2\sigma_{\%}$). The time is given in hours relative to the start of the PN or MOS1 exposures. The right y-axes give the relative differences in per cent between the quiet levels and the data. The upper x-axes show the stellar rotation phase, starting from the beginning of the observations. Error bars are the errors returned by the *epiclccorr* task. Small dark gray symbols visible in some panels indicate the background light curve in the same energy band. No simultaneous photometric observations are available for the event shown in the left column. For the other two events, simultaneous photometric fast mode observations in the UVW2 (middle) and UVM2 (right) bands are shown in the lowest panels. The fast mode data were rebinned to 200s. Bin widths are indicated by the horizontal bars, count rate errors were determined by standard error propagation of the errors returned by the *omfchain* task for the standard 10s binning. The dimming event in the middle column is only significant with the optimized binning method.

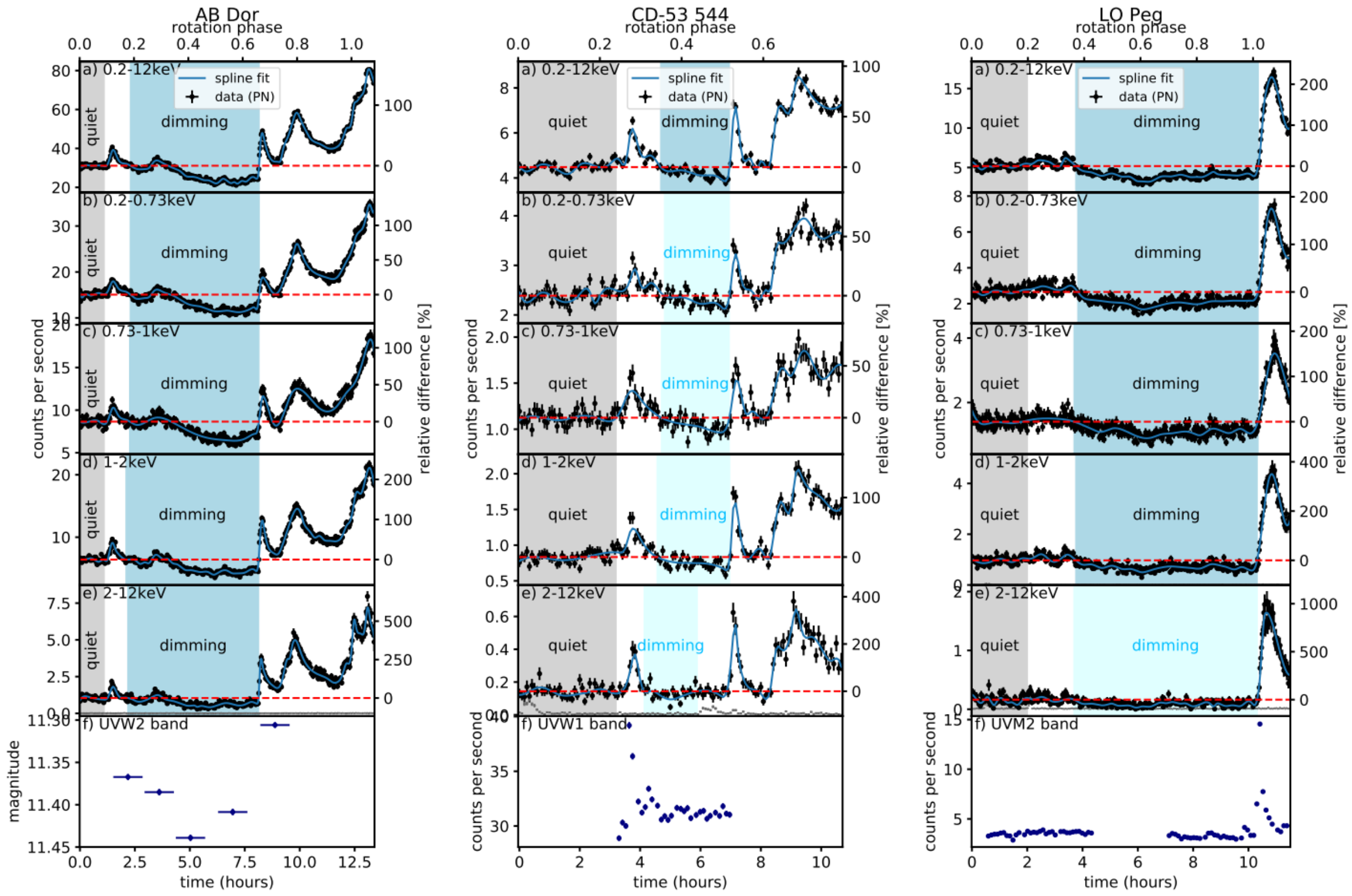

**Extended Data Figure 4**: Background subtracted X-ray light curves of the fast-rotating K-type stars AB Dor (left column), CD-53 544 (middle column) and LO Peg (right column). Data from the XMM EPIC PN detector are shown as filled symbols, and are plotted for the total energy band (0.2-12 keV) as well as four subbands. Bin sizes of 100s, 300s, and 100s were used (from left to right). The weighted spline fits to the data are shown as blue lines, the adopted quiet levels as red dashed lines. The quiet time intervals are indicated as gray shaded areas, the detected dimming regions as blue shaded areas (dark blue: maximum dimming depth >$2\sigma_{\%}$, light blue: maximum dimming depth <$2\sigma_{\%}$). The time is given in hours relative to the start of the PN exposures. The right y-axes give the relative differences in per cent between the quiet levels and the data. The upper x-axes show the stellar rotation phases, starting from the beginning of the observations. Error bars are the errors returned by the *epiclccorr* task. Small dark gray symbols visible in some panels indicate the background light curve in the same energy band. Simultaneous photometric imaging mode observations in the UVW2 bands (left), as well as fast mode observations in the UVW1 (middle; data in other bands before and after the event are omitted due to their different flux level) and UVM2 (right) bands are shown in the lowest panels. The fast mode data were rebinned to 400s. In the left column, exposure times are indicated by the horizontal bars, magnitude errors are the errors returned by the *omichain* task. In the other columns, bin widths are indicated by the horizontal bars, count rate errors were determined by standard error propagation of the errors returned by the *omfchain* task for the standard 10s binning.

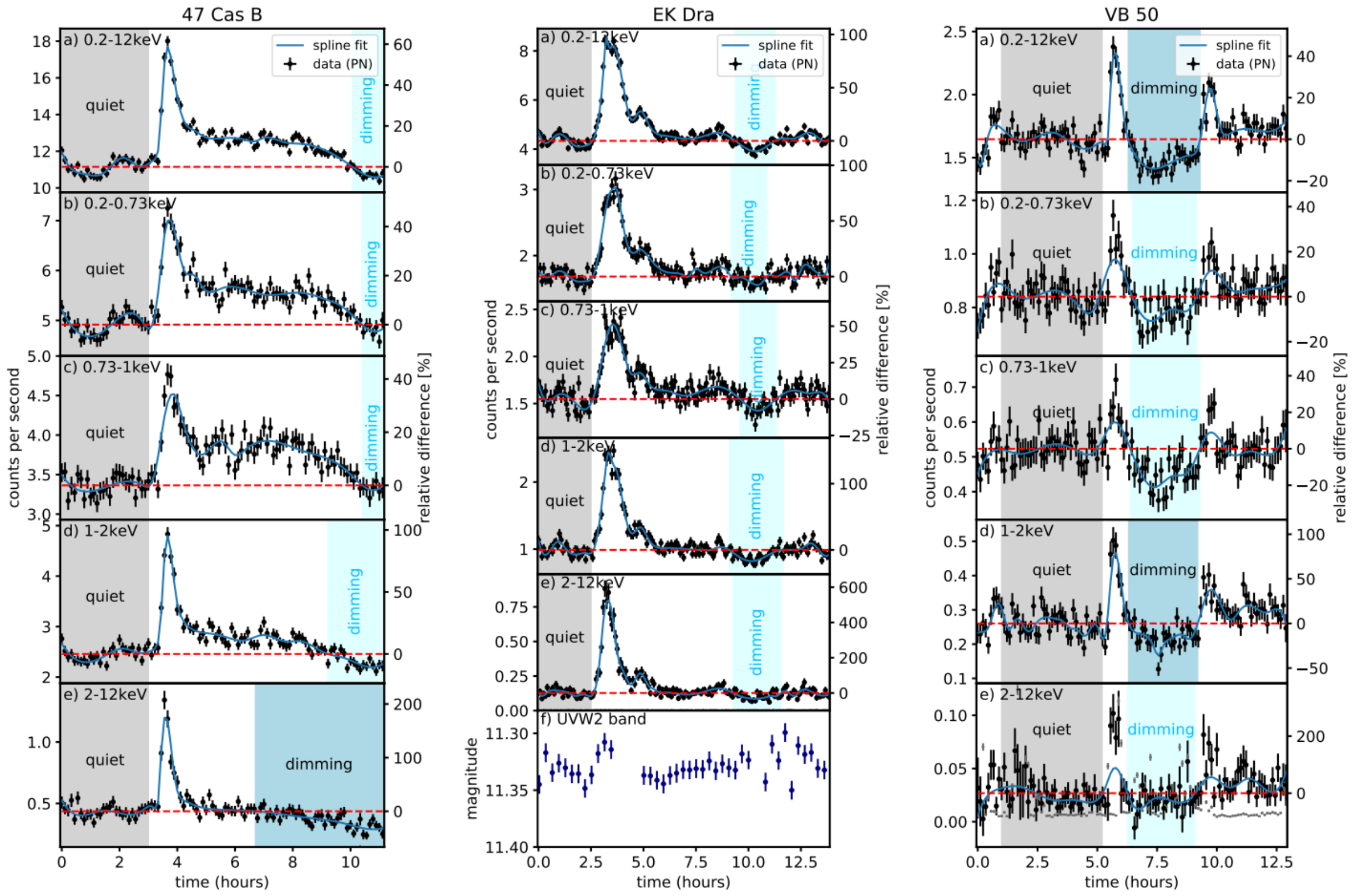

**Extended Data Figure 5:** Background subtracted X-ray light curves of the G-type stars 47 Cas B (left column), EK Dra (middle column) and VB 50 (right column). Data from the XMM EPIC PN detector are shown as filled symbols, and are plotted for the total energy band (0.2-12 keV) as well as four subbands. Bin sizes of 400s, 300s, and 400s were used (from left to right). The weighted spline fits to the data are shown as blue lines, the adopted quiet levels as red dashed lines. The quiet time intervals are indicated as gray shaded areas, the detected dimming regions as blue shaded areas (dark blue: maximum dimming depth >2$\sigma_{\%}$, light blue: maximum dimming depth <2$\sigma_{\%}$). The time is given in hours relative to the start of the PN exposures. The right y-axes give the relative differences in percent between the quiet levels and the data. Error bars are the errors returned by the *epiclccorr* task. Small dark gray symbols visible in some panels indicate the background light curve in the same energy band. No simultaneous photometric observations are available for 47 Cas B and VB 50. For EK Dra, simultaneous photometric imaging mode observations in the UVW2 band are shown in the lowest panel. Exposure times are indicated by the horizontal bars, magnitude errors are the errors returned by the *omichain* task. The dimming event of EK Dra is only significant (maximum dimming depth >2$\sigma_{\%}$) with the optimized binning method.

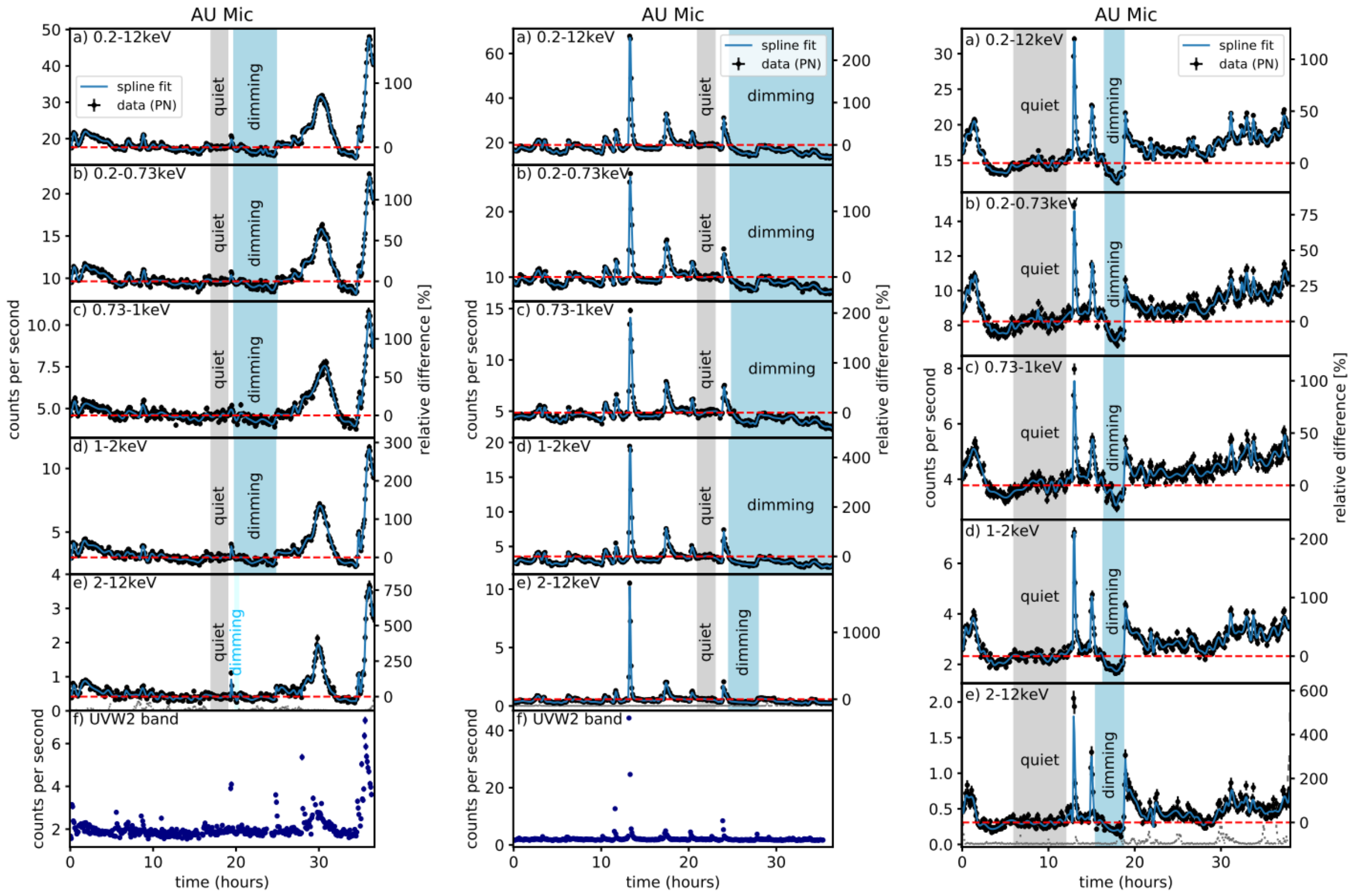

**Extended Data Figure 6**: Background subtracted X-ray light curves of the young M-type star AU Mic. Data from the XMM EPIC PN detector are shown as filled symbols, and are plotted for the total energy band (0.2-12 keV) as well as four subbands. Bin sizes of 300s were used. The weighted spline fits to the data are shown as blue lines, the adopted quiet levels as red dashed lines. The quiet time intervals are indicated as gray shaded areas, the detected dimming regions as blue shaded areas (dark blue: maximum dimming depth $>2\sigma_{\%}$, light blue: maximum dimming depth $<2\sigma_{\%}$). The time is given in hours relative to the start of the PN exposures. The right y-axes give the relative differences in per cent between the quiet levels and the data. Error bars are the errors returned by the *epiclccorr* task. Small dark gray symbols visible in some panels indicate the background light curve in the same energy band. Simultaneous photometric fast mode observations for the event shown in the right column are omitted, as they are corrupted after the first 7 hours of the observation. For the other two events, simultaneous photometric fast mode observations in the UVW2 band are shown in the lowest panels. The fast mode data were rebinned to 300s. Bin widths are indicated by the horizontal bars, count rate errors were determined by standard error propagation of the errors returned by the *omfchain* task for the standard 10s binning.

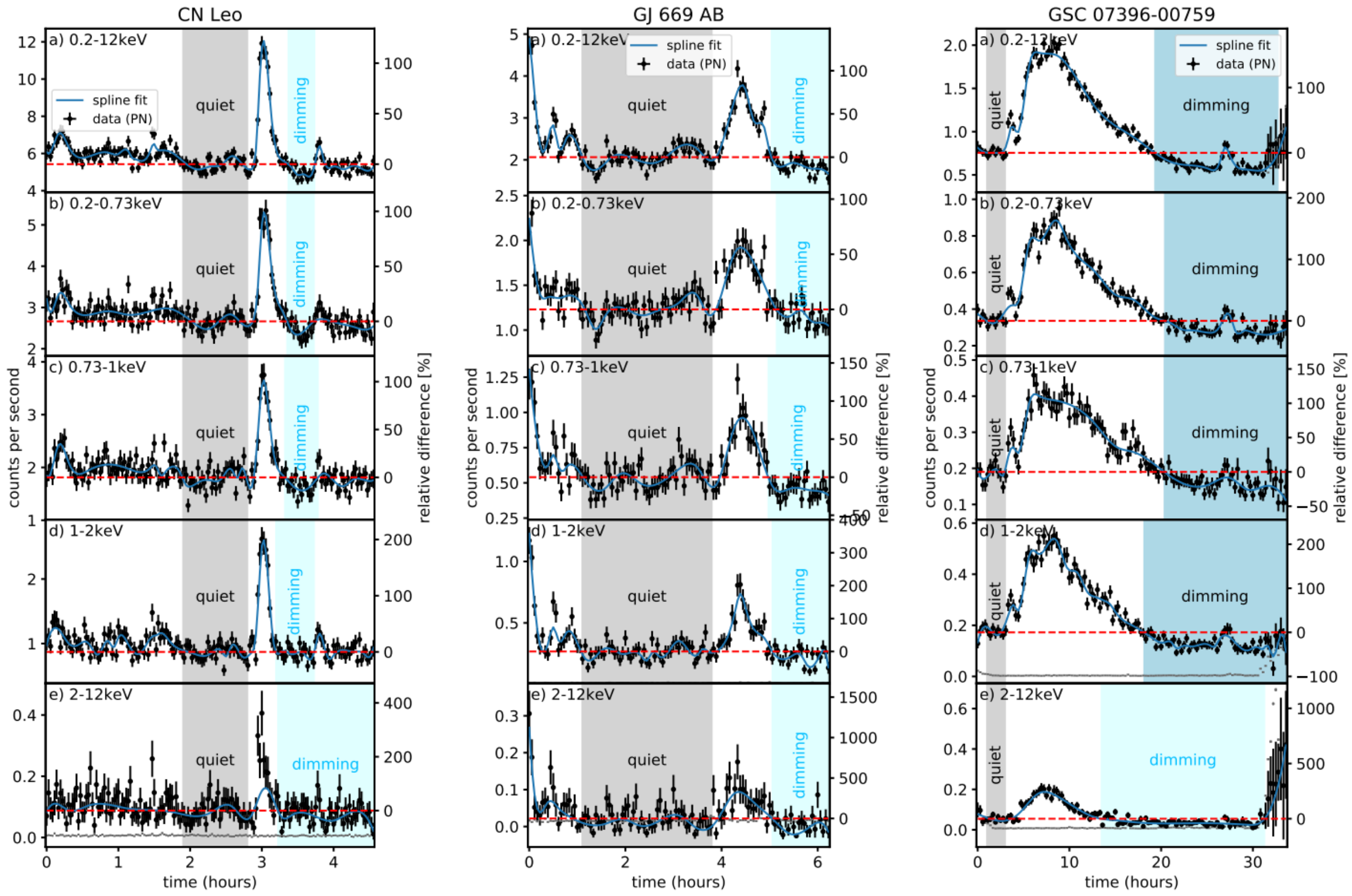

**Extended Data Figure 7**: Background subtracted X-ray light curves of the M-type stars CN Leo (left column), GJ 669 AB (middle column) and GSC 07396-00759 (right column). Data from the XMM EPIC PN detector are shown as filled symbols, and are plotted for the total energy band (0.2-12 keV) as well as four subbands. Bin sizes of 100s, 200s, and 1000s were used (from left to right). The weighted spline fits to the data are shown as blue lines, the adopted quiet levels as red dashed lines. The quiet time intervals are indicated as gray shaded areas, the detected dimming regions as blue shaded areas (dark blue: maximum dimming depth >2$\sigma_{\%}$, light blue: maximum dimming depth <2$\sigma_{\%}$). The time is given in hours relative to the start of the PN exposures. The right y-axes give the relative differences in per cent between the quiet levels and the data. Error bars are the errors returned by the *epiclccorr* task. Small dark gray symbols visible in some panels indicate the background light curve in the same energy band. For GJ 669 AB, the available imaging photometry consists of one image in each filter band, which cannot be used to create a photometric light curve. For CN Leo, they were omitted, as the fast mode observations are corrupted and the imaging mode observations have a too low cadence. For GSC 07396-00759, the imaging mode observations are also omitted because of their too low cadence. Moreover, the last few hours of this observation (time>30h) are affected by a strong background flare, making the data in the highest energy bin (lowest panel) unusable. The dimming events of CN Leo and GJ 669 AB are only significant with the optimized binning method.

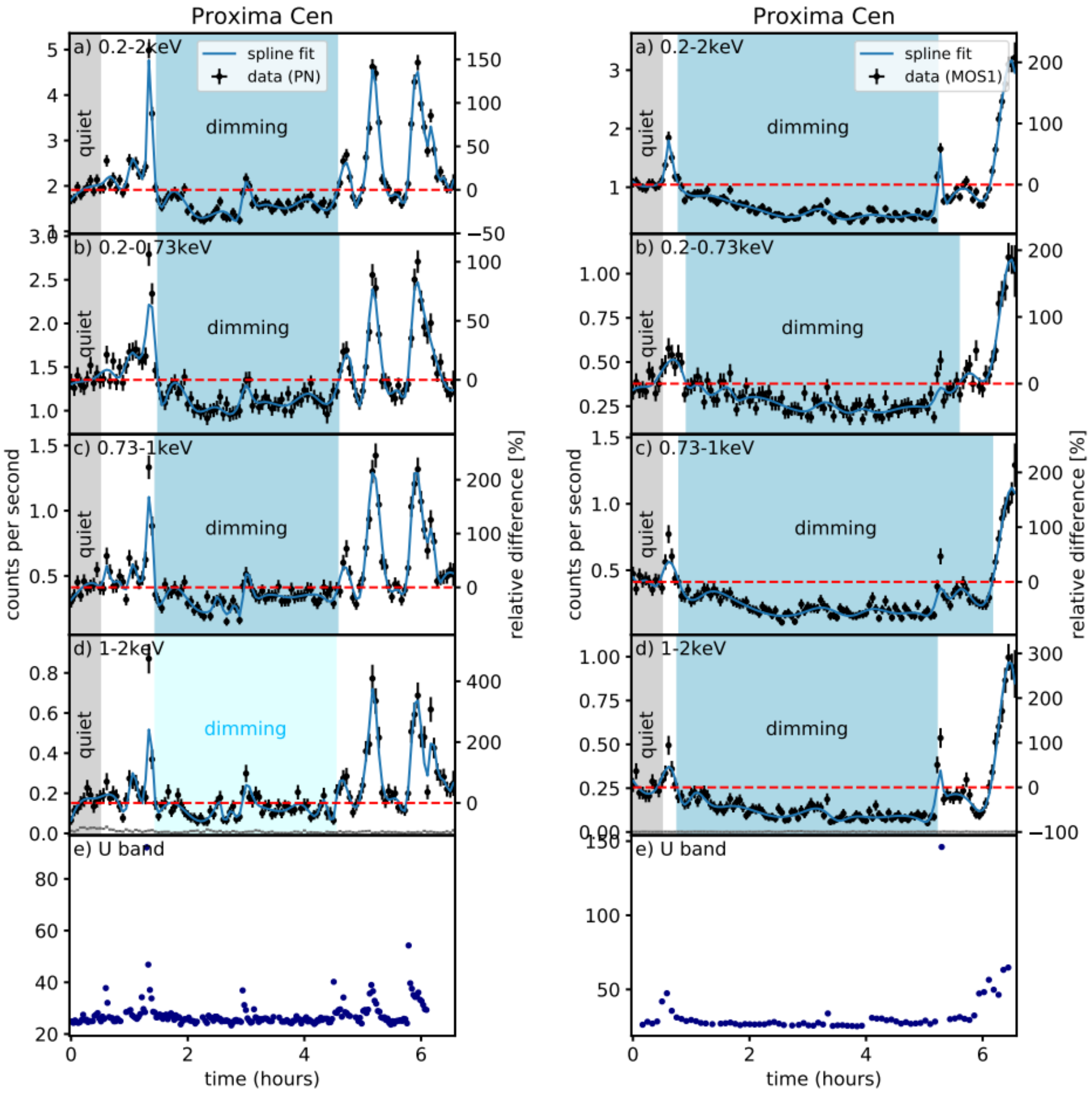

**Extended Data Figure 8:** Background subtracted X-ray light curves of Proxima Cen. Data from the XMM EPIC PN (left) and MOS1 (right) detectors are shown as filled symbols, and are plotted for the energy band 0.2-2 keV as well as four subbands (in both cases, data from 2-12 keV are not usable). Bin sizes of 200s and 400s were used (from left to right). The weighted spline fits to the data are shown as blue lines, the adopted quiet levels as red dashed lines. The quiet time intervals are indicated as gray shaded areas, the detected dimming regions as blue shaded areas (dark blue: maximum dimming depth >$2\sigma_{\%}$, light blue: maximum dimming depth <$2\sigma_{\%}$). The time is given in hours relative to the start of the PN or MOS1 exposures. The right y-axes give the relative differences in per cent between the quiet levels and the data. Error bars are the errors returned by the *epiclccorr* task. Small dark gray symbols visible in some panels indicate the background light curve in the same energy band. Simultaneous photometric fast mode observations in the U band are shown in the lowest panels. The fast mode data were rebinned to 100s (left) and 300s (right). Bin widths are indicated by the horizontal bars, count rate errors were determined by standard error propagation of the errors returned by the *omfchain* task for the standard 10s binning.

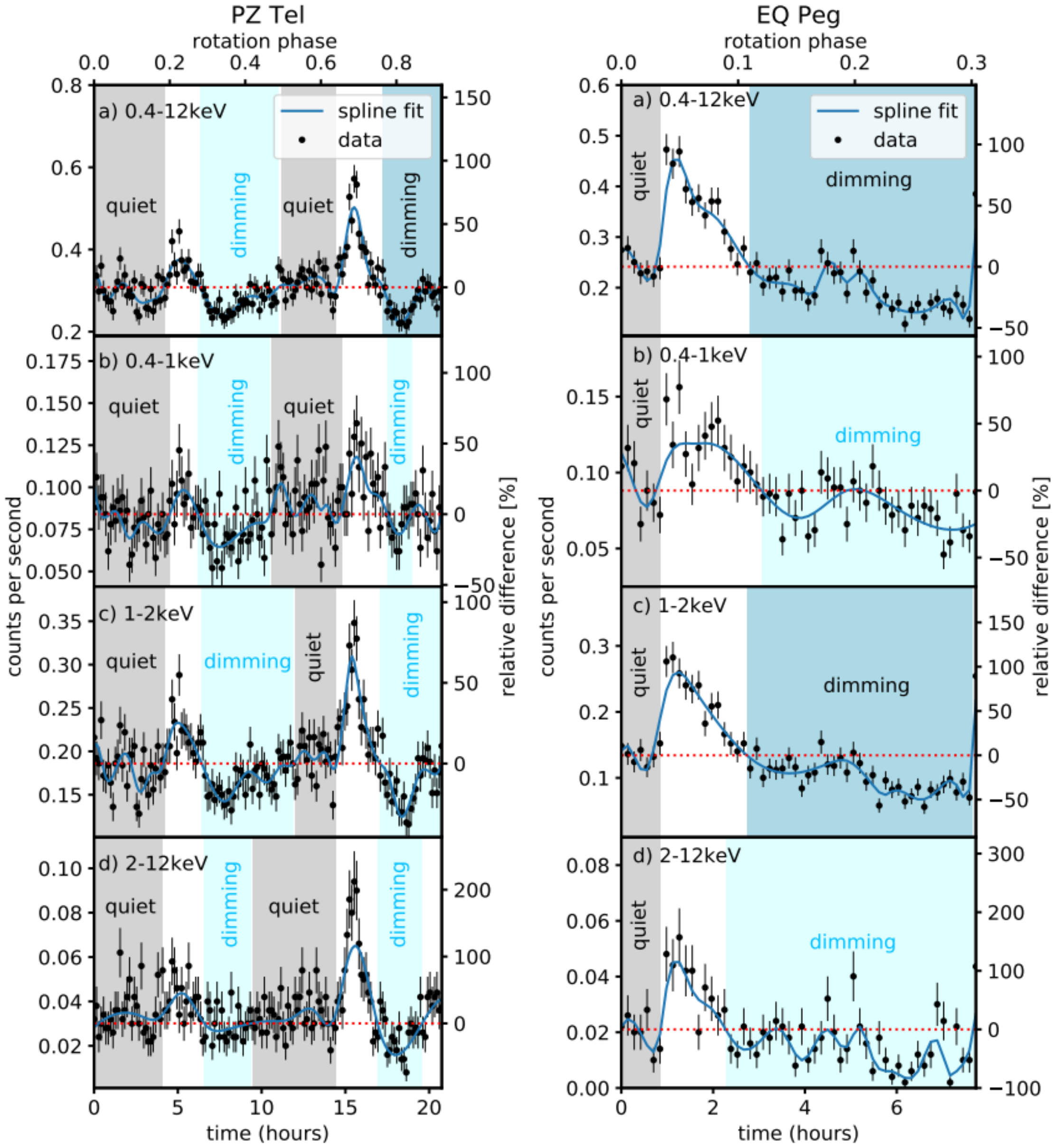

**Extended Data Figure 9**: Chandra ACIS HETGS light curves of the rapidly rotating young star PZ Tel (left panels) and the young binary star EQ Peg (right panels). The data are plotted for the total energy band (0.4-12 keV) as well as three subbands. All light curves are binned to 500s. The weighted spline fits to the data are shown as blue lines, the adopted quiet levels as red dashed lines. The quiet time intervals are indicated as gray shaded areas, the detected dimming regions as blue shaded areas (dark blue: maximum dimming depth >2$\sigma_{\%}$, light blue: maximum dimming depth <2$\sigma_{\%}$). The right y-axes give the relative differences in per cent between the quiet levels and the data. The upper x-axes show the stellar rotation phases, starting from the beginning of the observations. Error bars are the errors returned by the *aglc* task. The first dimming event of PZ Tel is only significant with the optimized binning method.

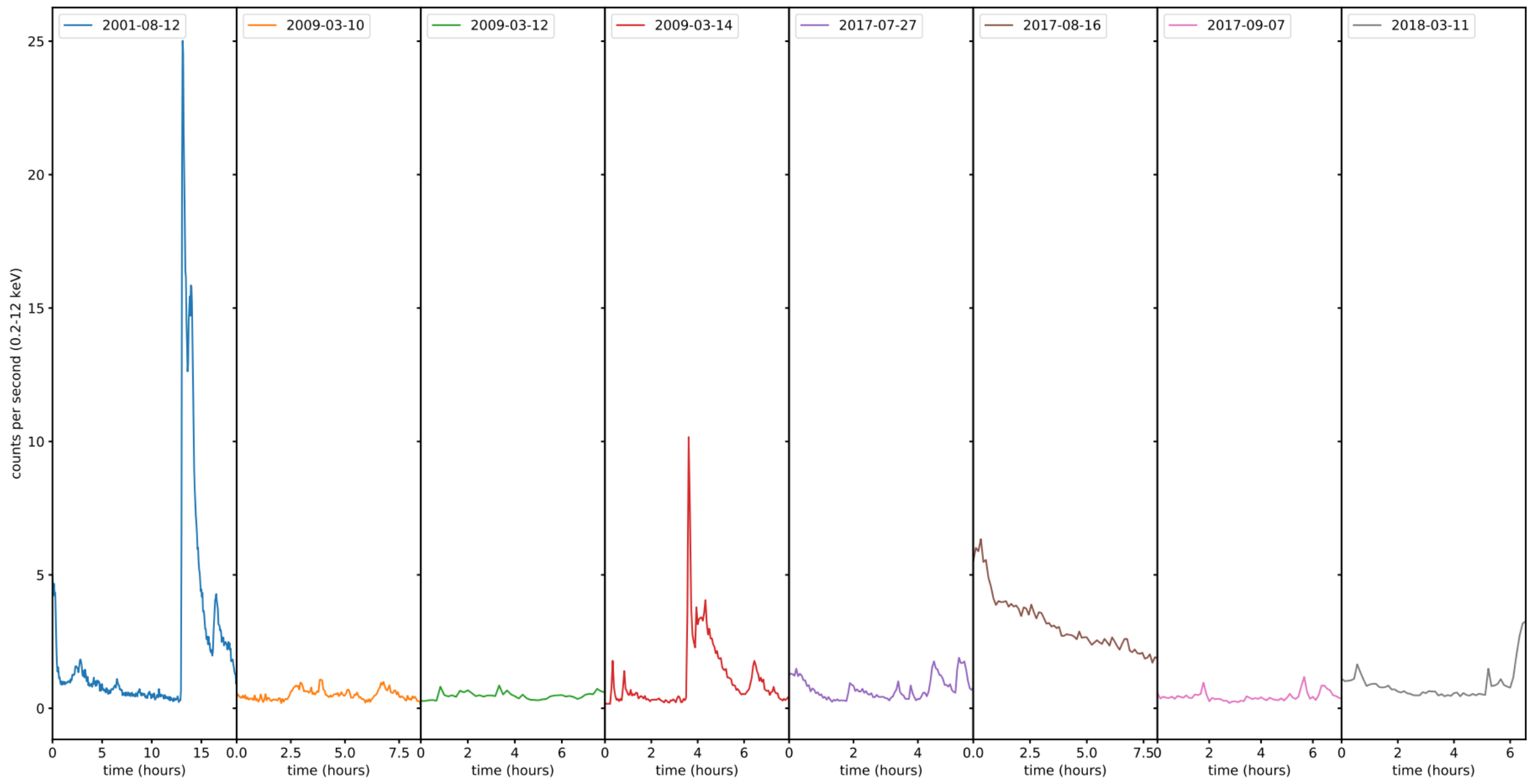

**Extended Data Figure 10**: Background subtracted X-ray (0.2-12 keV) light curves of Proxima Cen. This plot includes all eight available XMM observations with EPIC data, spanning the years from 2001 to 2018. Here we use the MOS1 exposures, because of their slightly longer durations than the PN exposures of the same observations. The dimming events are included in the last two observations. Error bars are omitted here for clarity. This representation shows that the dimming events are associated with small short flares, which can be understood in terms of better observability of the dimming compared to the bright flare emission.

# Supplementary Information

## Indications of stellar coronal mass ejections through coronal dimmings

Astrid M. Veronig[1,2], Petra Odert[1], Martin Leitzinger[1], Karin Dissauer[1,3], Nikolaus C. Fleck[1], Hugh S. Hudson[4,5]

1 Institute of Physics, University of Graz, Universitätsplatz 5, 8010 Graz, Austria
2 Kanzelhöhe Observatory for Solar and Environmental Research, University of Graz, Kanzelhöhe 19, 9521 Treffen, Austria
3 NorthWest Research Associates, 3380 Mitchell Lane, Boulder, CO 80301, USA
4 School of Physics & Astronomy, University of Glasgow, Glasgow G12 8QQ, UK
5 Space Sciences Laboratory, University of California, Berkeley, CA 94720, USA

In this supplementary file, we provide additional information on the individual stars, on which we identified coronal dimmings indicative of stellar CMEs. Supplementary Tables 1 and 2 list the main properties derived for the solar and stellar dimmings.

**47 Cas B**: This young Sun-like star is the secondary component of the close (0.1”) binary system 47 Cas, whose primary is an F0V star. The system could possibly include a third companion[61]. The X-ray emission of the system is attributed to the G star[61], as an F0 star at the given age (~100 Myr, Pleiades moving group member[62]) is unlikely to emit such a high flux. The flare analyzed here was already discussed in previous studies[40,63,64]. The XMM flare light curve in Extended Data Fig. 5 shows an extended plateau after an initial drop from its peak, with varying duration at different energy bands. No OM observations were available.

**AB Dor:** This young, fast-rotating, and very active K dwarf is the primary of a multiple system consisting of a close M dwarf binary at a separation of ~10”, as well as a close (~0.15”) brown dwarf[65]. It is a member of the AB Dor moving group with an age of ~150 Myr[66]. The K and M dwarf components are resolved in the XMM-Newton observations, but the close brown dwarf is not[65]. However, due to the high activity of the K star, it is not expected that the brown dwarf contributes a considerable amount of the measured X-ray flux of the system. AB Dor is the star with the largest number of dimming detections in our sample (five), and almost all light curves show several, often superimposed flare events (Fig. 3; Extended Data Figs. 3, 4). Due to the fast rotation (0.514 days[67]) of AB Dor, one observation typically covers more than one of its rotations. Rotational modulation is not detected in the EUV[45], whereas in soft X-rays the results are ambiguous. Some studies suggest the existence of rotational modulation of 5-13% (from ROSAT [0.6-12.4nm][68]) and of 12% (from Chandra [0.1-3.1nm][48]), while others find no significant detection (from XMM [0.1-6nm][69]; from ASCA [0.1-4nm][70]; from BeppoSAX [0.3-12.4 and 0.15-0.7nm][71]). As it seems, the determination of rotational modulation in soft X-rays on AB Dor depends on the time and the duration of the observations, as well as on the energy band. The soft X-ray light curves of AB Dor shown in Extended Data Figs. 3, 4 have durations of <1.5 rotation periods as well as numerous superimposed flares, and are therefore not suited to reliably deduce rotational modulations. The data shown in Extended Data Fig. 3 (right panel) was previously analyzed along with simultaneous optical spectroscopy[69]. In this study, it was excluded that the X-ray modulations are related to crossings of AB Dor’s large slingshot prominences. For the EUVE event on AB Dor shown in Fig. 3, rotational effects causing the dimming can be excluded, because the quiet level determined over several rotation periods did not show evidence for significant rotational modulation. These EUVE data of AB Dor were also studied previously[45,55].

**AU Mic:** This is a very young (~24 Myr, member of the β Pic moving group[66]) single M dwarf, which shows a common proper motion companionship with the widely separated M dwarf binary system AT Mic. Due to its young age, it still hosts a dusty debris disk, which is seen edge-on[72]. The star shows a high level of activity with many flares, and we detect dimming signatures after three of the flares (Extended Data Fig. 6). In this case, it is very unlikely that these dimming signatures are related to rotational modulation, as the rotation period is 4.85 days[67]. However, we note that the high flare occurrence frequency makes the determination of a quiet pre-flare level difficult, and dimming phases are often interrupted by subsequent flares. Fast mode OM observations turned out to be unusable for the event shown in the right column of Extended Data Fig. 6, as the star was not properly placed on the CCD chip after a long data gap.

**CD-53 544:** This active, fast rotating K dwarf is a member of a quadruple system (Extended Data Fig. 4, middle). It is a binary itself and hosts the M dwarf binary AF Hor as a companion at a separation of 22 arcsec[73]. It is a member of the Tuc Hor association with an age of ~45 Myr[66]. The star and its system bear some similarity to AB Dor. There are two rotation periods found in the literature for this star, 0.55976 days[74] and 1.275 days[67]. It is not uncommon that integer multiples of the rotation period are recovered in photometric rotation period searches. A quick visual inspection of the TESS light curves indicates a dominating sinusoidal variation with a period of about 0.5 days, therefore we adopt the former value. In a previous analysis of these observations, separate light curves of the visual binary components were constructed[75].

**CN Leo:** This is a single, active late-type M dwarf, rather similar to Proxima Cen in both spectral type and X-ray emission. Its rotation period is 2.7 days, as obtained from K2[76], although its activity level seems to be rather similar to the much slower rotating Proxima Cen. This could, however, be due to the fact that these stars are both fully convective. The flare event analyzed here (Extended Data Fig. 7, left column) was already presented in a previous study[77]. The fast mode OM imaging data of the observation turned out to be unusable (no star can be identified on the chip), and the imaging data does not have a sufficiently high cadence to provide additional constraints on the dimming event.

**EK Dra:** This young solar analog star has a close M dwarf companion[78], but due to its young age (~100 Myr, Pleiades moving group member[62]) and high activity level we assume that the G star dominates the X-ray emission of the system. The flare was analyzed in previous studies[79,40]. The XMM X-ray light curve (Extended Data Fig. 5, middle column) shows oscillatory behavior, which has been interpreted as quasi-periodic pulsations[80]. Their model of the event would include the dimming region identified here. The XMM OM data shows variability similar to the X-rays, but it is not clear whether the UV modulation also shows some dimming or not. The UV light curve generally increases during the dimming, except for one data point following a data gap shortly after the dimming maximum (Extended Data Fig. 5, panel f).

**EQ Peg:** The visual binary EQ Peg consists of two main-sequence stars of spectral type M3.5 and M4.5, both components known to be flaring stars[81,82]. It is a probable member of the Castor moving group[83] with an estimated age of 200 Myr. The Chandra data used in the present study have been analyzed previously, separately for each component of the EQ Peg binary system[82]. The light curves of EQ Peg shown in Extended Data Fig. 9 (right column) include both components. Those reveal one flare followed by a dimming. The quiescent level can be determined from a few data bins only.

**GJ 669 AB:** This young (0.6 Gyr, Hyades supercluster member[83]) M dwarf binary system has a separation of ~16 arcsec[73] and is not resolved by XMM-Newton, although clearly two components of the source can be seen in the X-ray images. The OM data consists of one or two images in each available filter of the field, therefore no light curve can be generated from these data. The XMM light curve of this event is shown in Extended Data Fig. 7, middle column.

**GSC 07396-00759:** This M-type weak line T-Tauri star with a debris disk is a physical companion to the classical T-Tauri star V4046 Sgr, a K dwarf binary system[84]. The stars are likely members of the β Pic moving group (~24 Myr[66]). The XMM observations show a large, long-duration flare (~15 hours) followed by a dimming (Extended Data Fig. 7, right column). The flare was previously analyzed[85]. The last few hours of the observations are contaminated by high background. Photometric fast mode OM observations are unavailable for this star, as they were only obtained for V4046 Sgr, the main target of the observation. The imaging photometry clearly shows the flare, but has a too low cadence to aid in the dimming analysis.

**LO Peg:** This is a single fast rotating K dwarf (0.423 days[67]), rather similar to AB Dor. The light curve morphology also shows some similarities (Extended Data Fig. 4, right column). The XMM OM data have a gap during a significant fraction of the dimming phase. The present observation has been analyzed previously, where the light curve shape was explained with rotational modulation[86]. However, the observations are only marginally longer than one period and affected by flares, which makes a robust determination of rotational modulation effects difficult.

**Proxima Cen**: Our nearest neighbor is an active late-type M dwarf and a widely separated companion to the α Cen system. Two planets have been detected, one of them in the habitable zone[87,88]. Proxima Cen shows two dimming events (Fig. 2, Extended Data Fig. 8), both occurring after rather small flares with energies of $4x10^{29}$ and $4.5x10^{29}$ erg in the 0.2-2 keV band, respectively. One dimming detection on Proxima Cen shows the largest depth of all stellar events, with a drop by 56%. Both dimmings are interrupted by subsequent flares due to the star's high flaring frequency. The events suffer from rather short coverage of the pre-flare phases, but both light curves are morphologically very similar to solar dimming events.

**PZ Tel:** This is a fast rotating (0.997 days[67]) solar analog pre-main sequence star. The age of PZ Tel is ~24 Myr[66], and its spectral type lies in the range of G6.5[89] to K8[67]. Prominences have been detected[90] and later confirmed[91]. The data have been already presented in a previous study[92]. The light curve shown in Extended Data Fig. 9 (left panel) reveals two flares each followed by dimmings.

**VB 50:** This is a Sun-like member of the Hyades cluster[93] (~600 Myr) and a close binary[94]. XMM light curves are shown in Extended Data Fig. 5 (right panel). There are no XMM OM observations available for this star.

**Solar dimming parameters and related flares.**

| Event | Flare Class | Flare Pos. [°] | CME | Dimm. EVE | Dimm. AIA | Dimming Depth [%] | $\sigma$ | $t_{dim}$ [h] | $t_{rise}$ [h] | $t_{rec}$ [h] | $t_{fp-dm}$ [h] | $t_{fp-ds}$ [h] |
|---|---|---|---|---|---|---|---|---|---|---|---|---|
| 20110213_1728 | M6.6 | S20 E04 | Y | Y | Y | 0.96±0.20 | 4.6 | 8.4 | 1.0 | 7.4 | 1.6 | 0.6 |
| 20110215_0144 | X2.2 | S20 W10 | Y | N | N | _ | _ | _ | _ | _ | _ | _ |
| 20110309_2313 | X1.5 | N08 W09 | N | N | N | _ | _ | _ | _ | _ | _ | _ |
| 20110730_0204 | M9.3 | N14 E35 | N | N | N | _ | _ | _ | _ | _ | _ | _ |
| 20110803_1317 | M6.0 | N16 W30 | Y | Y | Y | 2.29±0.19 | 12.5 | >9.7 | 2.5 | >7.2 | 3.3 | 0.8 |
| 20110804_0341 | M9.3 | N19 W36 | Y | Y | Y | 2.97±0.43 | 7.2 | >10.2 | 1.4 | >8.8 | 1.9 | 0.5 |
| 20110809_0748 | X6.9 | N17 W69 | Y | Y | Y | 2.30±0.18 | 13.4 | 9.4 | 3.9 | 5.5 | 4.6 | 0.7 |
| 20110906_0153 | M5.3 | N14 W07 | Y | Y | Y | 1.90±0.19 | 10.3 | 4.1 | 0.7 | 3.4 | 1.3 | 0.5 |
| 20110906_2212 | X2.1 | N14 W18 | Y | Y | Y | 3.86±0.20 | 20.5 | >10.5 | 0.5 | >10.0 | 0.8 | 0.3 |
| 20110907_2232 | X1.8 | N14 W28 | Y | Y | Y | 1.72±0.20 | 8.7 | >8.9 | 1.1 | >7.8 | 3.0 | 1.9 |
| 20110908_1532 | M6.7 | N14 W40 | Y | Y | Y | 0.38±0.18 | 2.1 | 1.2 | 0.3 | 0.9 | 0.9 | 0.6 |
| 20111103_2016 | X1.9 | N22 E63 | N | N | N | _ | _ | _ | _ | _ | _ | _ |
| 20120123_0338 | M8.7 | N28 W21 | Y | Y | Y | 0.77±0.20 | 3.8 | 4.1 | 1.2 | 2.9 | 3.6 | 2.4 |
| 20120127_1737 | X1.7 | N27 W71 | Y | Y | Y | 3.36±0.16 | 21.3 | >9.0 | 1.6 | >7.4 | 2.4 | 0.8 |
| 20120305_0230 | X1.1 | N17 E52 | Y | N | N | _ | _ | _ | _ | _ | _ | _ |
| 20120307_0002 | X5.4 | N17 E31 | Y | Y | Y | 5.50±0.23 | 25.2 | >7.2 | 1.6 | >5.6 | 3.1 | 1.5 |
| 20120309_0322 | M6.3 | N15 W03 | Y | Y | Y | 0.65±0.20 | 3.1 | >6.3 | 0.6 | >5.7 | 1.7 | 1.1 |
| 20120310_1715 | M8.4 | N17 W24 | Y | Y | Y | 0.65±0.28 | 2.3 | 2.2 | 1.3 | 0.9 | 3.7 | 2.4 |
| 20120510_0411 | M5.7 | N10 E22 | N | N | N | _ | _ | _ | _ | _ | _ | _ |
| 20120702_1043 | M5.6 | S17 E08 | Y | N | N | _ | _ | _ | _ | _ | _ | _ |
| 20120706_2301 | X1.1 | S13 W59 | Y | Y | Y | 1.90±0.27 | 7.1 | >9.9 | 2.1 | >7.8 | 3.0 | 0.9 |
| 20120712_1537 | X1.4 | S15 W01 | Y | Y | Y | 2.74±0.19 | 14.8 | >8.1 | 2.4 | >5.7 | 4.1 | 1.7 |
| 20130411_0655 | M6.5 | N09 E12 | Y | Y | Y | 1.01±0.18 | 5.8 | 3.4 | 2.4 | 1.0 | 3.3 | 0.9 |
| 20130513_0153 | X1.7 | N11 E89 | Y | Y | Y | 1.49±0.28 | 5.4 | 4.0 | 2.0 | 2.0 | 2.8 | 0.8 |
| 20130513_1548 | X2.8 | N11 E85 | Y | Y | Y | 1.08±0.19 | 5.7 | 4.1 | 1.3 | 2.8 | 3.0 | 1.7 |
| 20130514_0000 | X3.2 | N12 E67 | Y | Y | Y | 2.17±0.20 | 11.0 | >9.8 | 3.7 | >6.1 | 4.7 | 1.0 |
| 20130515_0125 | X1.2 | N12 E64 | Y | Y | Y | 1.16±0.16 | 7.4 | >9.3 | 1.2 | >8.1 | 2.4 | 1.2 |
| 20131024_0020 | M9.3 | S09 E10 | Y | Y | Y | 3.23±0.46 | 7.3 | 6.0 | 1.1 | 4.9 | 1.8 | 0.7 |
| 20131025_0753 | X1.7 | S08 E59 | Y | N | N | _ | _ | _ | _ | _ | _ | _ |
| 20131025_1451 | X2.1 | S08 E59 | Y | N | N | _ | _ | _ | _ | _ | _ | _ |
| 20131028_0141 | X1.0 | N04 W66 | Y | Y | Y | 3.10±0.49 | 6.5 | 2.1 | 1.3 | 0.8 | 1.8 | 0.5 |
| 20131029_2142 | X2.3 | N06 W91 | Y | Y | Y | 2.23±0.20 | 11.2 | >10.0 | 1.9 | >8.1 | 2.6 | 0.7 |
| 20131101_1946 | M6.3 | S12 E01 | N | Y | Y | 0.62±0.20 | 3.2 | 1.4 | 1.0 | 0.4 | 1.3 | 0.3 |
| 20131103_0516 | M4.9 | S12 W17 | N | N | Y | _ | _ | _ | _ | _ | _ | _ |
| 20131105_2207 | X3.3 | S12 E44 | Y | Y | Y | 1.84±0.23 | 7.7 | 4.2 | 0.6 | 3.6 | 0.8 | 0.2 |
| 20131108_0420 | X1.1 | S12 E13 | Y | Y | Y | 1.77±0.19 | 9.7 | 4.1 | 0.6 | 3.5 | 0.8 | 0.2 |
| 20131110_0508 | X1.1 | S14 W13 | Y | Y | Y | 1.96±0.20 | 9.7 | 3.8 | 0.7 | 3.1 | 1.0 | 0.3 |
| 20131119_1014 | X1.0 | S13 W79 | Y | Y | Y | 3.49±0.20 | 18.2 | >10.5 | 1.6 | >8.9 | 1.8 | 0.2 |
| 20131231_2145 | M6.4 | S15 W36 | Y | N | Y | _ | _ | _ | _ | _ | _ | _ |
| 20140107_1804 | X1.2 | S12 W08 | Y | Y | Y | 2.49±0.23 | 11.2 | >8.8 | 1.5 | >7.3 | 2.9 | 1.4 |
| 20140225_0039 | X4.9 | S12 E82 | Y | Y | Y | 3.58±0.25 | 15.1 | 8.7 | 2.2 | 6.5 | 2.8 | 0.6 |
| 20140329_1735 | X1.0 | N10 W32 | Y | Y | Y | 2.25±0.19 | 12.4 | >10.4 | 6.2 | >4.2 | 6.5 | 0.3 |
| 20140418_1231 | M7.3 | S20 W34 | Y | Y | Y | 2.11±0.20 | 10.5 | >9.3 | 2.5 | >6.8 | 3.7 | 1.2 |
| 20140425_0017 | X1.3 | S17 W91 | Y | Y | Y | 2.81±0.25 | 11.8 | >10.5 | 1.3 | >9.2 | 1.5 | 0.2 |
| 20110922_1029 | X1.4 | N11 E60 | Y | | N | | | | | | | |
| 20110924_0921 | X1.9 | N12 E60 | Y | | Y | | | | | | | |
| 20121023_0313 | X1.8 | S10E42 | N | | N | | | | | | | |
| 20140610_1136 | X2.2 | S15E80 | Y | | Y | | | | | | | |
| 20140610_1236 | X1.5 | S17E82 | Y | | Y | | | | | | | |
| 20140611_0859 | X1.0 | S18E65 | Y | | N | | | | | | | |
| 20140910_1659 | X1.6 | N11E05 | Y | | Y | | | | | | | |
| 20141019_0417 | X1.1 | S13E42 | N | | N | | | | | | | |
| 20141022_1402 | X1.6 | S14E13 | N | | N | | | | | | | |
| 20141022_0116 | M8.7 | S12E21 | N | | N | | | | | | | |
| 20141024_2050 | X3.1 | S22W21 | N | | N | | | | | | | |
| 20141025_1655 | X1.0 | S10W22 | N | | N | | | | | | | |
| 20141026_1004 | X2.0 | S14W37 | N | | N | | | | | | | |
| 20141027_1412 | X2.0 | S17W52 | N | | N | | | | | | | |
| 20141107_1653 | X1.6 | N17E40 | Y | | N | | | | | | | |
| 20141204_1805 | M6.1 | S20W31 | N | | N | | | | | | | |
| 20141217_0423 | M8.7 | S18E08 | Y | | N | | | | | | | |
| 20141220_0011 | X1.8 | S19W29 | Y | | N | | | | | | | |
| 20150309_2329 | M5.8 | S19E46 | Y | | Y | | | | | | | |
| 20150310_0319 | M5.1 | S15E39 | Y | | N | | | | | | | |
| 20150311_1611 | X2.1 | S12E22 | Y | | Y | | | | | | | |
| 20150622_1723 | M6.5 | N13W06 | Y | | N | | | | | | | |
| 20150625_0802 | M7.9 | N12W40 | Y | | N | | | | | | | |
| 20150928_1453 | M7.6 | S20W28 | N | | N | | | | | | | |

**Supplementary Table 1: Solar dimmings.** The dimmings are identified in full-Sun SDO/EVE 15-25 nm and SDO/AIA 19.3 nm light curves. Column 1 gives the date and start of the flare, column 2 the GOES soft X-ray flare class, column 3 the heliographic position of the flare. Column 4 states whether there was a CME associated with the flare. Columns 5 and 6 state whether a significant dimming was identified in the SDO/EVE 15-25 nm and AIA 19.3 nm light curves, respectively. Columns 7-11 give the derived SDO/EVE 15-25 nm dimming parameters: maximum depth and corresponding significance ($\sigma$), duration ($t_{dim}$), rise time ($t_{rise}$) and recovery time ($t_{rec}$). Columns 12 and 13 list the time between flare peak and dimming maximum ($t_{fp_dm}$) and the time between flare peak and dimming start ($t_{fp_ds}$), respectively.

## Stellar dimming parameters and properties of the stars.

| Star | Spectral type | $P_{rot}$ [d] | Age [Gyr] | Obs/DataID<br>Start time | Mission | Dimming Depth [%] | $\sigma$ | $t_{start}$ [h] | $t_{dim}$ [h] | $t_{rise}$ [h] | $t_{rec}$ [h] | $t_{fp_dm}$ [h] | $t_{fp_ds}$ [h] |
|---|---|---|---|---|---|---|---|---|---|---|---|---|---|
| 47 Cas B<br>Fig.E5/left | G0-2V | 1[62] | 0.1[62] | 0111520101<br>2001-09-11T02:21:19 | XMM | 5.04±3.82 | 1.3 | 10.1 | >1.0 | 0.8 | >0.2 | 7.2 | 6.4 |
| AB Dor<br>Fig.3 | K0V | 0.5[67] | 0.15[66] | ab_dor__9411121025N<br>1994-11-12T10:25:06 | EUVE | 33.73±11.94 | 2.8 | 95.6 | 11.4 | 1.2 | 10.2 | 3.1 | 1.9 |
| AB Dor<br>Fig.E3/left | | | | 0123720301<br>2000-10-27T15:23:55 | XMM | 12.23±3.31 | 3.7 | 4.7 | 1.4 | 1.1 | 0.3 | 4.2 | 3.1 |
| AB Dor<br>Fig.E3/middle | | | | 0412580701<br>2011-01-03T02:07:20 | XMM | 6.51±5.82 | 1.2 | 2.9 | >0.5 | 0.3 | >0.2 | 1.0 | 0.7 |
| AB Dor<br>Fig.E3/right | | | | 0602240201<br>2009-11-25T21:00:10 | XMM | 15.85±3.90 | 4.1 | 12.1 | 3.7 | 1.2 | 2.5 | 7.1 | 5.9 |
| AB Dor<br>Fig.E4/left | | | | 0134520701<br>2001-05-22T17:05:58 | XMM | 29.63±3.13 | 9.5 | 2.3 | 5.9 | 4.9 | 1.0 | 5.7 | 0.8 |
| AU Mic<br>Fig.E6/left | M0Ve | 4.85[67] | 0.024[66] | 0822740301<br>2018-10-10T13:13:52 | XMM | 11.94±2.34 | 5.1 | 19.7 | 5.1 | 4.7 | 0.4 | 5.0 | 0.3 |
| AU Mic<br>Fig.E6/middle | | | | 0822740401<br>2018-10-12T14:06:38 | XMM | 24.22±2.54 | 9.5 | 24.7 | >11.8 | 3.0 | >8.8 | 3.7 | 0.7 |
| AU Mic<br>Fig.E6/right | | | | 0822740501<br>2018-10-14T12:21:12 | XMM | 17.32±3.90 | 4.4 | 16.5 | 2.3 | 1.4 | 0.9 | 4.9 | 3.5 |
| CD-53 544<br>Fig.E4/middle | K6Ve | 0.56[74] | 0.045[66] | 0207210101<br>2004-02-09T21:27:52 | XMM | 14.13±5.76 | 2.5 | 4.7 | 2.3 | 2.2 | 0.1 | 3.1 | 0.9 |
| CN Leo<br>Fig.E7/left | M6Ve | 2.7[76] | <0.35[95] | 0200530301<br>2005-12-11T06:07:03 | XMM | 12.40±7.52 | 1.6 | 3.4 | 0.4 | 0.3 | 0.1 | 0.7 | 0.4 |
| EK Dra<br>Fig.E5/middle | G1.5V | 2.8[62] | 0.1[62] | 0111530101<br>2000-12-30T14:45:20 | XMM | 9.73±5.79 | 1.7 | 9.4 | 1.9 | 1.0 | 0.9 | 7.2 | 6.2 |
| EQ Peg<br>Fig.E9/right | M3.5Ve<br>+M4.5Ve | 1.1/<br>0.4[96] | 0.2[83] | 8486<br>2006-12-03T07:55:33 | Chandra | 42.26±13.62 | 3.1 | 2.8 | 4.9 | 4.7 | 0.2 | 6.2 | 1.5 |
| GJ 669 AB<br>Fig.E7/middle | M3.5Ve<br>+M4.5Ve | 20.5/<br>1.5[97] | 0.6[83] | 0500670301<br>2007-08-13T03:33:32 | XMM | 19.09±12.4 | 1.5 | 5.0 | >1.2 | 1.2 | >0 | 1.9 | 0.7 |
| GSC 07396-00759<br>Fig.E7/right | M1Ve | 12.1[84] | 0.024[66] | 0604860201<br>2009-09-15T01:21:00 | XMM | 27.41±9.92 | 2.8 | 19.3 | 13.4 | 11.2 | 2.2 | 22.2 | 11.0 |
| LO Peg<br>Fig.E4/right | K3Ve | 0.4[67] | 0.15[66] | 0740590101<br>2014-11-30T21:50:47 | XMM | 38.43±8.74 | 4.4 | 3.8 | 6.6 | 2.6 | 4.0 | 3.0 | 0.4 |
| Proxima Cen<br>Fig.2/Fig.E8/left | M5.5Ve | 83[98] | 4.85[99] | 0801880401<br>2017-09-07T07:44:08 | XMM | 35.57±9.37 | 3.8 | 1.5 | 3.1 | 1.4 | 1.7 | 1.5 | 0.1 |
| Proxima Cen<br>Fig.2/Fig.E8/right | | | | 0801880501<br>2018-03-11T20:01:51 | XMM | 56.45±8.63 | 6.5 | 0.8 | 4.5 | 2.9 | 1.6 | 3.1 | 0.2 |
| PZ Tel<br>Fig.E9/left | G9IV | 0.99[67] | 0.024[66] | 3729<br>2003-06-07T18:45:18 | Chandra | 22.73±13.14 | 1.7 | 6.3 | 4.7 | 1.5 | 3.2 | 2.7 | 1.2 |
| PZ Tel<br>Fig.E9/left | | | | 3729<br>2003-06-07T18:45:18 | Chandra | 28.45±12.70 | 2.2 | 17.2 | 3.5 | 1.1 | 2.4 | 2.8 | 1.7 |
| VB 50<br>Fig.E5/right | G1V | 7.1[93] | 0.6[93] | 0101440501<br>2002-03-05T17:47:46 | XMM | 14.22±6.06 | 2.4 | 6.3 | 3.0 | 1.0 | 2.0 | 1.6 | 0.6 |

**Supplementary Table 2: Stellar dimmings**. The first four columns give the characteristics of the star on which the dimming occurred (name, spectral type from SIMBAD[100], rotation period, age). In column 1, we also specify the corresponding plots shown in the Extended Data (indicated by an "E" before the Figure number) and the main text. Column 5 gives the unique observation IDs of the used data sets for each mission (column 6), as well as the start dates and times of the observations. Columns 7-12 give the derived dimming parameters: maximum depth and corresponding significance ($\sigma$), start time ($t_{start}$) of dimming relative to the start of the observation, duration ($t_{dim}$), rise time from dimming start to maximum ($t_{rise}$), and recovery time ($t_{rec}$). Columns 13 and 14 give the time between flare peak and dimming maximum ($t_{fp_dm}$) and the time between flare peak and dimming start ($t_{fp_ds}$), respectively. Note that not all dimming depths listed in column 7 are >2$\sigma_{\%}$, because the given values refer to the total energy band, but some dimmings are only significant in smaller subbands. The total energy bands are 0.2-12 keV for XMM-Newton (except for Proxima Cen, where only 0.2-2 keV were usable), 0.4-12 keV for Chandra, and 0.07-0.15 keV for EUVE.

**Other Supplementary Materials:** Supplementary Video 1

## References (Supplementary Information)